\documentclass{emulateapj}
\usepackage{apjfonts}

\newcommand{\Ha}{$\rm{H} \alpha$}
\newcommand{\Hb}{$\rm{H} \beta$}
\newcommand{\etal}{et~al.}
\newcommand{\PVdblt}{{\rm P}\kern 0.1em{\sc v}~$\lambda\lambda 1117, 1128$}
\newcommand{\CaIIdblt}{{\rm Ca}\kern 0.1em{\sc ii}~$\lambda\lambda 3934, 3969$}
\newcommand{\AlIIIdblt}{{\rm Al}\kern 0.1em{\sc iv}~$\lambda\lambda 1855, 1863$}
\newcommand{\CIVdblt}{{\rm C}\kern 0.1em{\sc iv}~$\lambda\lambda 1548, 1550$}
\newcommand{\MgIIdblt}{{\rm Mg}\kern 0.1em{\sc ii}~$\lambda\lambda 2796, 2803$}
\newcommand{\NVdblt}{{\rm N}\kern 0.1em{\sc v}~$\lambda\lambda 1238, 1242$}  
\newcommand{\SVIdblt}{{\rm S}\kern 0.1em{\sc vi}~$\lambda\lambda 933, 944$} 
\newcommand{\OVIdblt}{{\rm O}\kern 0.1em{\sc vi}~$\lambda\lambda 1031, 1037$} 
\newcommand{\SiIIdblt}{{\rm Si}\kern 0.1em{\sc ii}~$\lambda\lambda 1190, 1193$} 
\newcommand{\SiIVdblt}{{\rm Si}\kern 0.1em{\sc iv}~$\lambda\lambda 1393, 1402$} 
\newcommand{\PV}{\hbox{{\rm P}\kern 0.1em{\sc v}}}
\newcommand{\AlI}{\hbox{{\rm Al}\kern 0.1em{\sc i}}}
\newcommand{\AlII}{\hbox{{\rm Al}\kern 0.1em{\sc ii}}}
\newcommand{\AlIII}{{\hbox{\rm Al}\kern 0.1em{\sc iii}}}
\newcommand{\CaII}{\hbox{{\rm Ca}\kern 0.1em{\sc ii}}}
\newcommand{\CII}{\hbox{{\rm C}\kern 0.1em{\sc ii}}}
\newcommand{\CIIe}{\hbox{{\rm C$^{\ast}$}\kern 0.1em{\sc ii}}}
\newcommand{\CIII}{\hbox{{\rm C}\kern 0.1em{\sc iii}}}
\newcommand{\CIV}{\hbox{{\rm C}\kern 0.1em{\sc iv}}}
\newcommand{\CV}{\hbox{{\rm C}\kern 0.1em{\sc v}}}
\newcommand{\HI}{\hbox{{\rm H}\kern 0.1em{\sc i}}}
\newcommand{\HII}{\hbox{{\rm H}\kern 0.1em{\sc ii}}}
\newcommand{\Lya}{\hbox{{\rm Ly}\kern 0.1em$\alpha$}}
\newcommand{\Lyb}{\hbox{{\rm Ly}\kern 0.1em$\beta$}}
\newcommand{\Lyg}{\hbox{{\rm Ly}\kern 0.1em$\gamma$}}
\newcommand{\Lyd}{\hbox{{\rm Ly}\kern 0.1em$\delta$}}
\newcommand{\Lye}{\hbox{{\rm Ly}\kern 0.1em$\epsilon$}}
\newcommand{\Lyphi}{\hbox{{\rm Ly}\kern 0.1em$\phi$}}
\newcommand{\Lyfive}{\hbox{{\rm Ly}\kern 0.1em$5$}}
\newcommand{\Lysix}{\hbox{{\rm Ly}\kern 0.1em$6$}}
\newcommand{\Lyseven}{\hbox{{\rm Ly}\kern 0.1em$7$}}
\newcommand{\Lyeight}{\hbox{{\rm Ly}\kern 0.1em$8$}}
\newcommand{\Lynine}{\hbox{{\rm Ly}\kern 0.1em$9$}}
\newcommand{\Lyten}{\hbox{{\rm Ly}\kern 0.1em$10$}}
\newcommand{\Lyeleven}{\hbox{{\rm Ly}\kern 0.1em$11$}}
\newcommand{\HeI}{\hbox{{\rm He}\kern 0.1em{\sc i}}}
\newcommand{\HeII}{\hbox{{\rm He}\kern 0.1em{\sc ii}}}
\newcommand{\FeI}{\hbox{{\rm Fe}\kern 0.1em{\sc i}}}
\newcommand{\FeII}{\hbox{{\rm Fe}\kern 0.1em{\sc ii}}}
\newcommand{\FeIII}{\hbox{{\rm Fe}\kern 0.1em{\sc iii}}}
\newcommand{\MnII}{\hbox{{\rm Mn}\kern 0.1em{\sc ii}}}
\newcommand{\MgI}{\hbox{{\rm Mg}\kern 0.1em{\sc i}}}
\newcommand{\MgIb}{\hbox{{\rm Mg}\kern 0.1em{\sc i}}\kern 0.05em{\rm b}}
\newcommand{\MgII}{\hbox{{\rm Mg}\kern 0.1em{\sc ii}}}
\newcommand{\MgIII}{\hbox{{\rm Mg}\kern 0.1em{\sc iii}}}
\newcommand{\NI}{\hbox{{\rm N}\kern 0.1em{\sc i}}}
\newcommand{\NII}{\hbox{{\rm N}\kern 0.1em{\sc ii}}}
\newcommand{\NIII}{\hbox{{\rm N}\kern 0.1em{\sc iii}}}
\newcommand{\NV}{\hbox{{\rm N}\kern 0.1em{\sc v}}}
\newcommand{\OVI}{\hbox{{\rm O}\kern 0.1em{\sc vi}}}
\newcommand{\OI}{\hbox{{\rm O}\kern 0.1em{\sc i}}}
\newcommand{\OII}{\hbox{[{\rm O}\kern 0.1em{\sc ii}]}}
\newcommand{\OIII}{\hbox{[{\rm O}\kern 0.1em{\sc iii}]}}
\newcommand{\OIV}{\hbox{{\rm O}\kern 0.1em{\sc iv}]}}
\newcommand{\SI}{{\rm S}\kern 0.1em{\sc i}}
\newcommand{\SIV}{{\rm S}\kern 0.1em{\sc iv}}
\newcommand{\SVI}{{\rm S}\kern 0.1em{\sc vi}}
\newcommand{\SiI}{\hbox{{\rm Si}\kern 0.1em{\sc i}}}
\newcommand{\SiII}{\hbox{{\rm Si}\kern 0.1em{\sc ii}}}
\newcommand{\SiIII}{\hbox{{\rm Si}\kern 0.1em{\sc iii}}}
\newcommand{\SiIV}{\hbox{{\rm Si}\kern 0.1em{\sc iv}}}
\newcommand{\SII}{\hbox{{\rm S}\kern 0.1em{\sc ii}}}
\newcommand{\SIII}{\hbox{{\rm S}\kern 0.1em{\sc iii}}}
\newcommand{\NaI}{\hbox{{\rm Na}\kern 0.1em{\sc i}}}
\newcommand{\NaID}{\hbox{{\rm Na}\kern 0.1em{\sc i}}\kern 0.05em{\rm D}}
\newcommand{\TiII}{\hbox{{\rm Ti}\kern 0.1em{\sc ii}}}
\newcommand{\kms}{\hbox{km~s$^{-1}$}}
\newcommand{\cmsq}{\hbox{cm$^{-2}$}}

\slugcomment{Accepted Feb. 20 2011}

\shorttitle{\sc Galaxy Disk and Halo Gas Kinematics at $z\sim0.1$}
\shortauthors{\sc Kacprzak {\etal}}

\begin{document}


\title{Halo Gas and Galaxy Disk Kinematics of a Volume-Limited Sample
of {\MgII} Absorption-Selected Galaxies at $z\sim0.1$}


\author{\sc
Glenn G. Kacprzak\altaffilmark{1},
Christopher W. Churchill\altaffilmark{2},
Elizabeth J. Barton\altaffilmark{3},
and
Jeff Cooke\altaffilmark{1}
}
\altaffiltext{1}{Swinburne University of Technology, Victoria 3122,
Australia; {\tt gkacprzak@astro.swin.edu.au, jcooke@astro.swin.edu.au}}

\altaffiltext{2}{Center for Cosmology, Department of Physics and
Astronomy, University of California, Irvine, CA 92697-4575, USA; {\tt
ebarton@uci.edu}}
 
\altaffiltext{3}{New Mexico State University, Las Cruces, NM 88003 USA;
{\tt cwc@nmsu.edu}}

\begin{abstract}

  We have directly compared {\MgII} halo gas kinematics to the
  rotation velocities derived from emission/absorption lines of the
  associated host galaxies.  Our $0.096 \leq z \leq 0.148$
  volume-limited sample comprises 13 $\sim L_{\star}$ galaxies, with
  impact parameters of $12-90$ kpc from background quasars
  sight-lines, associated with 11 {\MgII} absorption systems with
  {\MgII} equivalent widths $0.3 \leq W_r(2796) \leq 2.3$~{\AA}.  For
  only 5/13 galaxies, the absorption resides to one side of the galaxy
  systemic velocity and trends to align with one side of the galaxy
  rotation curve. The remainder have absorption that spans both sides
  of the galaxy systemic velocity.  These results differ from those at
  $z\sim 0.5$, where 74\% of the galaxies have absorption residing to
  one side of the galaxy systemic velocity.  For all the $z\sim 0.1$
  systems, simple extended disk-like rotation models fail to reproduce
  the full {\MgII} velocity spread, implying other dynamical processes
  contribute to the {\MgII} kinematics. In fact 55\% of the galaxies
  are ``counter-rotating'' with respect to the bulk of the {\MgII}
  absorption.  These {\MgII} host-galaxies are isolated, have low star
  formation rates (SFRs) in their central regions ($\lesssim
  1$~M$_{\odot}$~yr$^{-1}$), and SFRs per unit area well below those
  measured for galaxies with strong winds.  The galaxy {\NaID}
  (stellar$+$ISM) and {\MgIb} (stellar) absorption line ratios are
  consistent with a predominately stellar origin, implying
  kinematically quiescent interstellar media.  These facts suggest
  that the kinematics of the {\MgII} absorption halos for our sample
  of galaxies are not influenced by galaxy--galaxy environmental
  effects, nor by winds intrinsic to the host galaxies.  For these low
  redshift galaxies, we favor a scenario in which infalling gas
  accretion provides a gas reservoir for low-to-moderate star
  formation rates and disk/halo processes.

\end{abstract}




\section{Introduction}

Absorption lines detected in the spectra of background quasars and
gamma-ray bursts provide powerful probes of the intervening Universe.
In particular, it was predicted that discrete metal lines, produced in
the clumpy metal enriched gas distribution arising from intervening
groups and isolated galaxies, should be detected as absorption in
quasar spectra \citep{bahcall65,bahcall66,bahcall69}.  Since the
{\MgIIdblt} doublet arises in metal-enriched low ionization gas, with
neutral hydrogen column densities of $10^{16} \lesssim \hbox{N(\HI)}
\lesssim 10^{22}$~{\cmsq} \citep{archiveI,weakII}, it should be a
dominant absorption feature detected in quasar spectra.  In fact,
{\MgII} absorption was indeed detected along quasar sight-lines within
close proximity to, and similar redshifts as, foreground galaxies
\citep{bergeron88,bb91}.

Since then, the association of {\MgII} to normal, bright, field
galaxies is now well established \citep[see][ and references
therein]{cwc-china}. It has been demonstrated that galaxy {\MgII}
halos extend out to $\sim$120~kpc and have gas covering fractions of
50--80\% \citep{tripp-china,kacprzak08,chen08,chen10a}. However,
despite the recent progress on the galaxy-halo connection, the origins
of this extended halo gas is still widely debated.

Host galaxy environment has been demonstrated to play some role in
determining the gas covering fraction and extent of their halos from
studies of galaxy close pairs, galaxy groups and galaxy clusters
associated with {\MgII} absorption
\citep{nestor07,lopez08,padilla09,chen10a,kacprzak10a,kacprzak10b}.
Although the bulk of the evidence supports {\MgII} absorption detected
along quasar sight-lines is a result of outflowing gas
\citep[e.g.,][]{bond01a,ellison03,bouche06,bouche08,menard09,nestor10},
several studies are now emerging that support infalling gas as the
source of the absorption
\citep{chen08,chen10a,kacprzak10a,kacprzak10c,stocke10} or a
combination of both inflow and outflow \citep{chen10b,chelouche10}.
In fact, \citet{kacprzak10c} have suggested that different mechanisms
may be responsible for different equivalent width regimes: low
equivalent width systems ($\lesssim1$~\AA) may be dominated by infall
and the strong equivalent width systems may be dominated by outflows.

It is difficult to isolate separate dynamic processes, such as inflow
and outflow, responsible for producing the {\MgII} absorption by only
studying statistically large samples of absorbers. However, it may be
possible to further understand these individual dynamic processes by
studying galaxies--absorber pairs on case-by-case basis.  Furthermore,
studying a direct comparison of the galaxy disk kinematics {\it and\/}
absorbing {\MgII} halo gas kinematics may yield a more in depth
understanding of these processes.

A direct comparison of the galaxy disk kinematics and absorbing
{\MgII} halo gas kinematics has been performed for 19 $z\sim 0.6$
systems \citep{steidel02,ellison03,chen05,kacprzak10a}. The galaxy
halos were probed over a range of impact parameters, $8\leq D \leq
110$~kpc. \citet{kacprzak10a} studied 10 host galaxy/absorber pairs
and found that the absorption was fully to one side of the galaxy
systemic velocity and usually aligned with one arm of the rotation
curve in most cases.  These results are consistent with earlier
studies of five galaxies by \citet{steidel02}, one galaxy by
\citet{ellison03}, and three galaxies of \citet{chen05}.  In only 5/19
cases, the absorption velocities span both sides of the galaxy
systemic velocity. These observations are highly suggestive that
{\MgII} halos obey ``disk-like'' rotation dynamics, given the
alignment of the absorption and the the velocity offsets of the
absorbing gas relative to the galaxy.

\citet{steidel02} applied simple disk--halo models and concluded that a
large fraction of the {\MgII} halo gas velocities could be explained
by an extension of the disk rotation with some lag in velocity.
However, the models were not able to account for the full velocity
spread of the gas.  This was also confirmed by \citet{kacprzak10a}
where a large fraction of the {\MgII} could not be explained by disk
halo rotation alone and the observed additional velocities were
consistent with infalling gas as demonstrated by their study of
hydrodynamical galaxy simulations within a cosmological context.

With the increasing blue sensitivity of CCDs, a new redshift window
has recently opened, enabling us to detect photons down to the
atmospheric cutoff of around 3050~{\AA}. This provides a lower {\MgII}
absorption detection redshift of $z=0.09$. Only a few studies have
taking advantage of this new regime to explore the evolution of
covering fractions, halo sizes and kinematics as a function of
redshift \citep{barton09,chen10a,chen10b}.

We have obtained the rotation curves of 13 $\sim L_{\star}$ galaxies,
at $0.096 \leq z \leq 0.148$, that also have Keck/LRIS quasar
absorption profiles of {\MgII}.  In this paper we perform a kinematic
comparison of these galaxies and their associated halo {\MgII}
absorption. We compare our data with a simple rotating disk--halo
model, implemented in \citet{kacprzak10a} and \citet{steidel02}, and
examine the maximum absorption velocities that could be attributed to
the halo gas as disk rotation alone.  We further examine the host
galaxy environments and also study the intrinsic host galaxy
properties, such as SFRs, {\MgIb} and {\NaID} line ratios, and {\NaID}
and {\Ha} velocity offsets, that are used to identify and quantify
strong outflows.  The paper is organized as follows: In
\S~\ref{sec:datakine}, we present our sample, and explain the data
reduction and analysis. In \S~\ref{fields}, we present the results of
our galaxy--{\MgII} absorption kinematic observations, and in
\S~\ref{halo}, we compare the observed {\MgII} absorption velocities
with a simple disk kinematic halo model. In \S~\ref{sec:mech} we
evaluate and discuss the potential mechanisms that produce the
observed {\MgII} absorption detected near these galaxies. We end with
our discussion and conclusions in \S~\ref{sec:discussionkine} and
\S~\ref{sec:conclusionkine}, respectively.  Throughout we adopt an
H$_{\rm 0}=70$~\kms Mpc$^{-1}$, $\Omega_{\rm M}=0.3$,
$\Omega_{\Lambda}=0.7$ cosmology.

\begin{deluxetable*}{lcrrcrrl}
\tabletypesize{\scriptsize} \tablecaption{Keck--I/LRIS Quasar
Observations\label{tab:LRIS}} \tablecolumns{8} \tablewidth{0pt}

\tablehead{
\colhead{SDSS}&
\colhead{Foreground}&
\colhead{RA} &
\colhead{DEC} &
\colhead{Quasar } &
\colhead{ } &
\colhead{ } &
\colhead{Exposure}\\
\colhead{Quasar Name}&
\colhead{Galaxy Name} &
\colhead{(J2000)} &
\colhead{(J2000)} &
\colhead{$m_r$} &
\colhead{$z_{em}$} &
\colhead{Date (UT)} &
\colhead{(sec.)}
}
\startdata
SDSS J005244.23$-$005721.7 & J005244G1  & 00:52:44.23 & $-$00:57:21.82 & 18.8  &	 0.756 &  Sep. 18 2009  &$2\times940$\\
                           & J005244G2  &             &                &       &               &                &	\\
SDSS J111850.13$-$002100.7 & J111850G1  & 11:18:50.14 & $-$00:21:00.61 & 18.9  &	 1.026 &  Jan. 12 2010  &$1\times1245$\\
SDSS J114518.47$+$451601.4 & J114518G1  & 11:45:18.48 & $+$45:16:04.45 & 18.7  &         0.612 &  Jan. 12 2010  &$2\times1245$\\
SDSS J161940.56$+$254323.0 & J161940G1  & 16:19:40.57 & $+$25:43:23.12 & 16.8  &	 0.269 &  Jul. 21 2009  &$1\times600$\\
SDSS J225036.72$+$000759.4 & J225036G1  & 22:50:36.74 & $+$00:07:59.45 & 19.1  &	 0.431 &  Jul. 21 2009  &$1\times700$ 
\enddata
\end{deluxetable*}

\begin{deluxetable*}{lccccrcrl}
\tabletypesize{\scriptsize} \tablecaption{APO/DIS and Keck--II/ESI
Galaxy Observations\label{tab:DIS}} \tablecolumns{9} \tablewidth{0pt}

\tablehead{
\colhead{ }&
\colhead{RA} &
\colhead{DEC} &
\colhead{Galaxy} &
\colhead{Galaxy} &
\colhead{ } &
\colhead{Instrument} &
\colhead{Slit} &
\colhead{Exposure}\\
\colhead{Galaxy Name}&
\colhead{(J2000)} &
\colhead{(J2000)} &
\colhead{$m_r$} &
\colhead{$M_r$} &
\colhead{Date (UT)} &
\colhead{-- Telescope} &
\colhead{PA} &
\colhead{(sec.)}
}
\startdata
J005244G1  & 00:52:43.92  & $-$00:57:09.23& 16.8 & $-$20.8&  Nov. 16 2009&   DIS/APO &   35  & 3$\times$1600\\	   
J005244G2  & 00:52:44.02  & $-$00:56:46.41& 19.5 & $-$19.7&  Nov. 15 2009&   DIS/APO &$-$60  & 3$\times$1600\\	   
J081420G1  & 08:14:22.08  & $+$38:33:49.32& 16.7 & $-$20.1&  Dec. 01 2008&   DIS/APO &    0  & 3$\times$1700\\	   
J091119G1  & 09:11:16.80  & $+$03:12:10.69& 16.8 & $-$20.0&  Feb. 02 2009&   DIS/APO &   50  & 4$\times$1700\\	   
J092300G1  & 09:23:00.96  & $+$07:51:05.11& 16.4 & $-$20.6&  Mar. 02 2009&   DIS/APO &   22  & 3$\times$1800\\	   
J102847G1  & 10:28:46.56  & $+$39:18:42.78& 17.1 & $-$20.1&  May  08 2008&   DIS/APO &   94  & 3$\times$ 600\\	    
J111850G1  & 11:18:49.68  & $-$00:21:10.02& 17.1 & $-$20.4&  Jan. 18 2010&   DIS/APO &  126  & 3$\times$1800\\	   
J114518G1  & 11:45:19.92  & $+$45:16:10.56& 17.0 & $-$20.5&  Jan. 18 2010&   DIS/APO &   45  & 3$\times$1800\\	    
J114803G1  & 11:48:03.84  & $+$56:54:25.92& 16.5 & $-$20.5&  Dec. 01 2008&   DIS/APO &   30  & 3$\times$1200\\	   
J144033G1  & 14:40:35.52  & $+$04:48:50.43& 16.9 & $-$20.3&  May  08 2008&   DIS/APO &   50  & 4$\times$ 960\\	   
J144033G2  & 14:40:34.56  & $+$04:48:25.10& 17.2 & $-$20.2&  Aug. 07 2010&   ESI/Keck&  110  & 2$\times$ 500\\
J161940G1  & 16:19:39.36  & $+$25:43:33.60& 18.1 & $-$19.0&  Apr. 07 2010&   DIS/APO &  120  & 4$\times$1400\\
J225036G1  & 22:50:37.70  & $+$00:07:45.02& 15.9 & $-$21.2&  Nov. 16 2009&   DIS/APO &   60  & 3$\times$1600	   
\enddata
\end{deluxetable*}
\section{Data and Analysis}
\label{sec:datakine}

Our sample consists of 11 {\MgII} absorption systems detected in the
spectra of background quasars that are associated with 13 foreground
galaxies at similar redshifts. Six of the 11 {\MgII} absorption
systems were selected from \citet{barton09} who performed a
volume--limited survey targeting $\sim L_{\star}$ galaxies at
$z\sim0.1$ with $M_r + \mbox{log}h \gtrsim -20.5$ that were in close
proximity to quasar lines-of-sight. They detected six {\MgII}
absorption systems that are at a similar redshift to seven foreground
$\sim L_{\star}$ galaxies.  We are in the process of expanding this
survey to a more luminous absolute magnitude limit (-21) and, hence, a
somewhat higher redshift (Barton et al. 2011).  Ultimately, the study
will involve approximately 45 galaxies.  Here, we add an additional
five {\MgII} absorption systems, associated with six galaxies,
discovered from the expanded survey. Here, we focus on the first
kinematics study of the absorbing galaxies at $z\sim 0.1$. We do not
include the non-absorbing galaxies identified in this survey in this
study.  Thus, our sample is composed of 11 absorption systems that are
associated with 13 galaxies, two of which are double galaxy systems.
The galaxy--quasar impact parameters range from $12 \leq D \leq
90$~kpc.  We discuss our data and the analysis in the next
subsections.

\subsection{Quasar Spectroscopy}\label{sec:qso_spec}

In addition to the six quasar spectra obtained in \citet{barton09}, we
acquired an additional five quasar spectra between 2009 July 21 and
2010 January 12.  Details of the additional observations are presented
in Table~\ref{tab:LRIS}. We used the LRIS-B/Keck 1200 lines/mm
grating, blazed at 3400~{\AA}, which covers a wavelength range of
2910$-$3890\AA.  Using a 1.0$''$ slit results in a dispersion of
0.24~{\AA} per pixel and provides a resolution of $\sim$1.6~{\AA}
($\sim$150~\kms). Integration times of 600--2490 seconds were used,
depending on the magnitude of the quasar, providing 3$\sigma$
detection limits of $W_r(2796) \gtrsim 0.2$\AA.

The spectra were reduced using the standard IRAF
packages\footnote{IRAF is written and supported by the IRAF
  programming group at the National Optical Astronomy Observatories
  (NOAO) in Tucson, Arizona. NOAO is operated by the Association of
  Universities for Research in Astronomy (AURA), Inc.\ under
  cooperative agreement with the National Science Foundation.}. Since
neither the sky nor the quartz lamps provide substantial photon counts
at $\sim$3100\AA, the data were not flat fielded and no sky background
correction was applied. The spectra are heliocentric and vacuum
corrected.

The quasar continuum fit was obtained iteratively.  First, low order
orthonormal polynomials were fitted to the low frequency shape of the
photon counts over the full wavelength range, then higher frequency
features and emission lines were fit to localized wavelength regions
using multiple Gaussian functions and orthonormal polynomials
\citep[see][]{sembach92}.

The uncertainty spectrum was created post-reduction using a simple
Poisson (counts plus background), flat field, and a read noise model
appropriate for the sky conditions and instrument specifications.  A
small scale factor was applied to ensure a reduced $\chi^2$ about the
continuum fit of unity within a tolerance of 0.1 (iteratively
rejecting outlier pixels such as those associated with absorption
features).

The {\MgIIdblt} doublets were objectively searched for using the
methods described by \citet{weakI}.  Because of the low redshifts of
the target systems, the redshift number density of interloping
absorption lines in the spectra are negligible; as such, there was no
confusion with blends or misidentifications.  Significant (3~$\sigma$)
corroborating transitions such as {\MgI} $\lambda 2852$ and {\FeII}
$\lambda\lambda 2344,2383,2600$ were identified using the {\it a
posteriori\/} knowledge of the {\MgII} absorption redshift.

Analysis of the absorption profiles was performed using graphic-based
interactive software of our own design, which uses the direct pixel
values to measure the equivalent widths, velocity moments, and the
redshift of the {\MgII} $\lambda 2796$ transition.  Absorption system
velocity widths were measured between the pixels where the equivalent
width per resolution element recover to the $1~\sigma$ detection
threshold \citep{weakI}.  The redshift for each {\MgII} system is
computed from the optical depth weighted mean of the absorption
profile \citep[see][]{cv01}. The statistical uncertainties in the
redshifts range between 0.00001--0.00009 ($\sim 3$--$30$~{\kms}
co-moving).  In addition, least-squares Gaussian deblending was
performed \citep[using the program FITTER;][]{archiveI} to estimate the
equivalent widths, velocity widths, and velocity centroids of
component structures in the absorption profiles. We only display the
Gaussian fits in Figures \ref{fig:plotVr2}--\ref{fig:plotVr5} to help
the reader identify the {\MgII} absorption profiles and we only use
the direct pixel-measured values for all quantitative values publish
here.  Our equivalent widths published here differ from those of
\citet{barton09}, who choose to publish the Gaussian fitted equivalent
widths.


\begin{deluxetable*}{lccrrccccrr}
\tabletypesize{\scriptsize} \tablecaption{Galaxy Redshifts and
{\MgII} Absorption Properties\label{tab:zdat}}

\tablecolumns{11} \tablewidth{0pt}
\tablehead{
\colhead{Galaxy Name}&
\colhead{$D$} &
\colhead{$z_{gal}$} &
\colhead{$z_{abs}$} &
\colhead{$\Delta v_{r}$\tablenotemark{ a}}&
\colhead{$W_r(2796)$} &
\colhead{$W_r(2803)$} &
\colhead{DR\tablenotemark{ b} }&
\colhead{$\Delta V_-$}\tablenotemark{c} &
\colhead{$\Delta V_+$}\tablenotemark{c}\\
\colhead{ }&
\colhead{(kpc) }&
\colhead{ }&
\colhead{ }&
\colhead{(\kms)}&
\colhead{(\AA)} &
\colhead{(\AA)} &
\colhead{ } &
\colhead{(\kms) } &
\colhead{(\kms) } 
}
\startdata
J005244G1 & $86.1\pm1.2$ & 0.13429$\pm$0.00005 & 0.13460$\pm$0.00002& $-$14.7   & 1.46$\pm$0.04 &  1.23$\pm$0.04 &   1.19$\pm$0.05&  $-$353.8 &   241.9 \\ 
J005244G2 & $32.4\pm0.2$ & 0.13465$\pm$0.00002 &                    &  $+$82.2  &               &                &                &           &         \\
J081420G1 & $51.1\pm0.3$ & 0.09801$\pm$0.00004 & 0.09833$\pm$0.00001&  $+$105.2 & 0.57$\pm$0.05 &  0.28$\pm$0.05 &   2.04$\pm$0.37&      21.7 &   178.5 \\ 
J091119G1 & $72.1\pm0.4$ & 0.09616$\pm$0.00004 & 0.09636$\pm$0.00009&  $+$66.3  & 0.82$\pm$0.10 &  0.34$\pm$0.07 &   2.41$\pm$0.59&  $-$336.6 &   454.0 \\
J092300G1 & $11.9\pm0.3$ & 0.10385$\pm$0.00005 & 0.10423$\pm$0.00004&  $+$127.1 & 2.25$\pm$0.14 &  1.40$\pm$0.12 &   1.61$\pm$0.17&  $-$186.3 &   418.2 \\

J102847G1 & $89.8\pm0.4$ & 0.11348$\pm$0.00002 & 0.11411$\pm$0.00004&  $+$168.3 & 0.30$\pm$0.02 &  0.13$\pm$0.02 &   2.23$\pm$0.36&  $-$54.9  &   329.5 \\
J111850G1 & $25.1\pm0.3$ & 0.13159$\pm$0.00001 & 0.13158$\pm$0.00002&  $-$4.8   & 1.93$\pm$0.08 &  1.82$\pm$0.07 &   1.06$\pm$0.06&  $-$253.2 &   227.7 \\
J114518G1 & $39.4\pm0.8$ & 0.13389$\pm$0.00004 & 0.13402$\pm$0.00002&  $+$46.0  & 1.06$\pm$0.06 &  1.07$\pm$0.05 &   0.99$\pm$0.07&  $-$188.8 &   247.8 \\
J114803G1 & $29.1\pm0.5$ & 0.10451$\pm$0.00011 & 0.10433$\pm$0.00002&  $-$62.1  & 1.59$\pm$0.06 &  1.25$\pm$0.05 &   1.27$\pm$0.07&  $-$341.0 &   217.3 \\   
J144033G1 & $67.1\pm0.1$ & 0.11277$\pm$0.00005 & 0.11304$\pm$0.00001&  $+$90.7  & 1.18$\pm$0.04 &  0.93$\pm$0.03 &   1.28$\pm$0.06&  $-$115.0 &   293.4 \\   
J144033G2 & $24.9\pm0.2$ & 0.11271$\pm$0.00001 &                    &  $+$88.8  &               &                &                &           &         \\
J161940G1 & $45.7\pm0.7$ & 0.12438$\pm$0.00006 & 0.12501$\pm$0.00003&  $+$211.8 & 0.32$\pm$0.03 &  0.28$\pm$0.03 &   1.12$\pm$0.18&      72.5 &   370.9 \\    
J225036G1 & $53.9\pm0.7$ & 0.14826$\pm$0.00002 & 0.14837$\pm$0.00002&  $+$38.0  & 1.08$\pm$0.07 &  1.11$\pm$0.07 &   0.97$\pm$0.09&  $-$140.8 &   224.4      
\enddata
\tablenotetext{a}{$\Delta v_{r}$ is the rest--frame velocity offset
between the mean {\MgII} $\lambda 2976$ absorption line and the galaxy
where, $\Delta v_{r}=c(z_{abs}-z_{gal})/(1+z_{gal})$~\kms.}
\tablenotetext{b}{Doublet ratio, $DR\equiv W_r(2796)/W_r(2803)$.}
\tablenotetext{c}{Blue and red {\MgII} absorption velocity edges.}
\end{deluxetable*}

\subsection{Galaxy Spectroscopy}

The majority of the galaxy spectra were obtained during 9 nights
between 2008 May and 2010 February using the double imaging
spectrograph (DIS) at the Apache Point Observatory (APO) 3.5m
telescope in New Mexico. Details of the observations are presented in
Table~\ref{tab:DIS}. The total exposure time per target ranges from
1800 to 5400 seconds. For each galaxy, the slit position angle was
selected to lie along the galaxy major axis.

The DIS spectrograph has separate red and blue channels that have
plate scales of 0.40$''$~pixel$^{-1}$ and 0.42$''$~pixel$^{-1}$,
respectively.  The B1200 grating was used for the blue channel
resulting in a spectral resolution of 0.62~\AA~pixel$^{-1}$ with
wavelength coverage of 1240~\AA. The R1200 grating was used for the
red channel resulting in a spectral resolution of
0.58~\AA~pixel$^{-1}$ with wavelength coverage of 1160~\AA.
Wavelength centers for each grating were selected to target $z\sim0.1$
galaxies having either $\rm{H} \alpha$ plus [\NII] (red channel) and
{\OII} (blue channel) in emission or [\NaI] (red channel) and {\CaII}
H \& K (blue channel) in absorption. We used a 1.5$''$-wide by
6$'$-long slit with no on-chip binning of the CCD.  The spectral
resolution is 1.76~{\AA} ($\sim$120~\kms) and 1.28~{\AA}
($\sim$53~\kms) FWHM in the blue and red channels, respectively.  The
observations were performed during good weather conditions with
typical seeing of 1$-$2$''$.

The spectrum of galaxy J144033G2 was obtained using the Keck Echelle
Spectrograph and Imager, ESI, \citep{sheinis02} on 2010 August 07.
Details of the ESI/Keck observations are presented in
Table~\ref{tab:DIS}. The slit length is $20''$ and $1''$ wide and we
used $2\times1$ on-chip CCD binning in the spatial direction. Binning
by two in the spatial directions results in pixel sizes of
$0.27-0.34''$ over the echelle orders of interest. The mean seeing was
$0.8''$ (${\rm FWHM}$) with clear skies. The total exposure time was
1000s.  The wavelength coverage of ESI is 4000 to 10,000~{\AA}, which
allow us to obtain multiple emission lines (such as {\OII} doublet,
$\rm{H}\beta$, {\OIII} doublet, $\rm{H}\alpha$, [\NII] doublet, etc.)
with a velocity dispersion of $11$~\kms~pixel$^{-1}$ (${\rm
  FWHM}\sim45$~km/s).

All DIS and ESI galaxy spectra were reduced using IRAF.  External
quartz dome-illuminated flat fields were used to eliminate
pixel-to-pixel sensitivity variations.  Each DIS spectrum was
wavelength calibrated using HeNeAr arc line lamps, and the ESI
spectrum was calibrated using CuArXe arc line lamps.  Again, galaxy
spectra are both vacuum and heliocentric velocity corrected for
comparison with the absorption line kinematics.  The arc emission line
vacuum wavelengths were obtained from the National Institute of
Standards and Technology (NIST) database.

A Gaussian fitting algorithm \citep[FITTER: see][]{archiveI}, which
computes best fit Gaussian amplitudes, line centers, and widths, was
used to obtain the galaxy redshifts from an emission or absorption
line(s).  The galaxy redshifts derived here are consistent with those
derived by SDSS.  The galaxy redshifts are listed in
Table~\ref{tab:zdat}; their accuracy ranges from 2--30~\kms.

The rotation curve extraction was performed following the methods of
\citet{kacprzak10a} \citep[also see][]{vogt96,steidel02}. We extracted
individual spectra by summing three-pixel-wide apertures
(corresponding to approximately one resolution element of
$1.20-1.26''$ for DIS and $0.81-1.02''$ for ESI) at one pixel spatial
increments along the slit.  To obtain accurate wavelength
calibrations, we extract spectra of the arc line lamps at the same
spatial pixels as the extracted galaxy spectra.  Fitted arc lamp
exposures provided a dispersion solution accurate to $\sim0.15$~{\AA}
($6$~{\kms}) and $\sim0.025$~{\AA} ($1$~{\kms}) for DIS and ESI,
respectively. Each galaxy emission line (or absorption line in some
cases) was fit with a single Gaussian in order to extract the
wavelength centroid for each line.  The velocity offsets for each
emission line in each extraction were computed with respect to the
redshift zero point determined for the galaxy (see
Table~\ref{tab:zdat}).  The rotation curves for the 13 galaxies are
presented in Figures \ref{fig:plotVr2}--\ref{fig:plotVr5}.

\subsection{Galaxy Images \& Models}

\begin{figure*}
\begin{center}
\includegraphics[angle=90,scale=0.40]{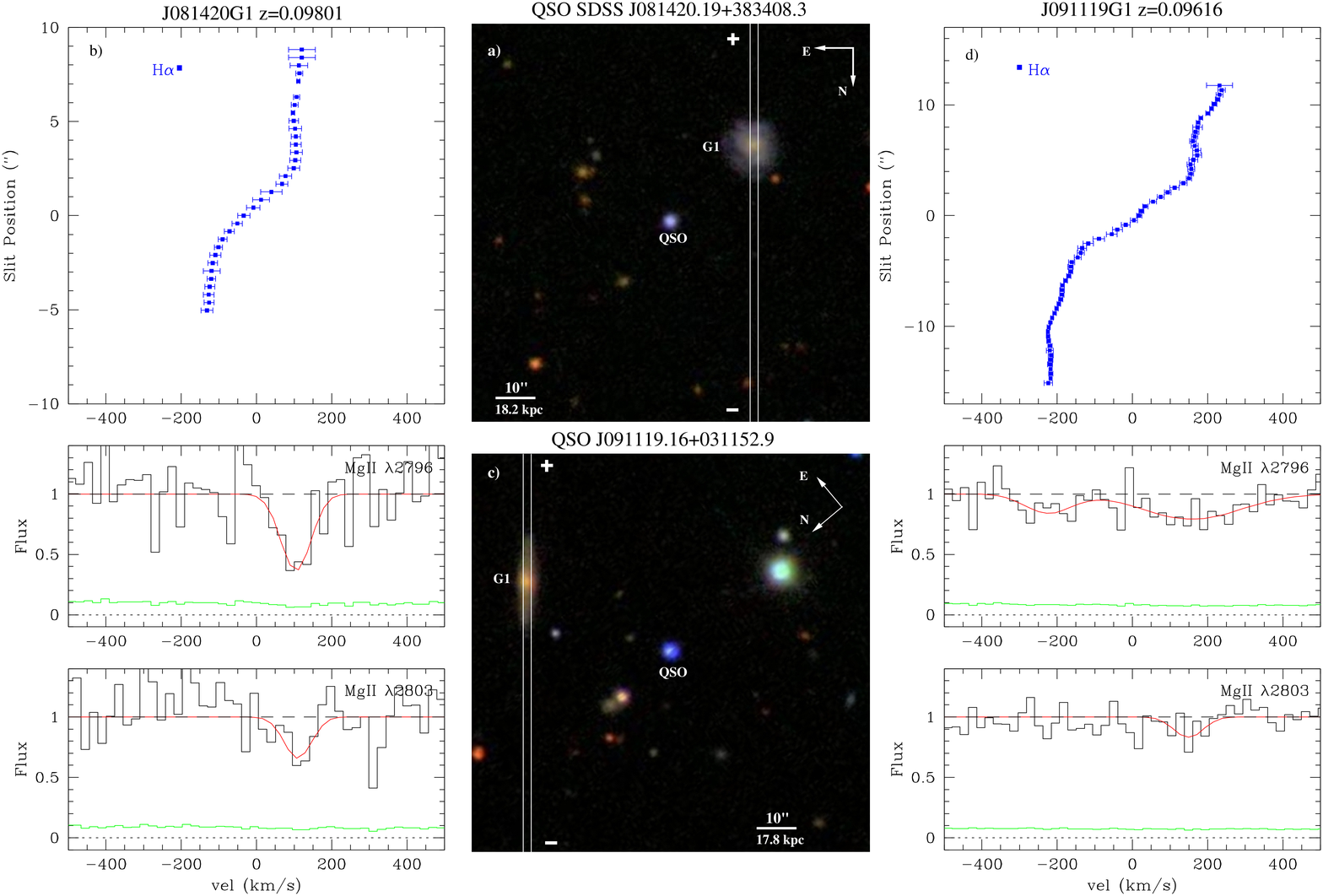}
\caption{(a) A $100'' \times 100''$ (181.8 $\times$ 181.8~kpc at the
  absorption redshift) Sloan $gri$ color image of the SDSS
  J081420.19+383408.3 quasar field. The DIS/APO slit is superimposed
  on the image over the absorbing galaxy J081420G1.  The '$+$' and
  '$-$' on the slit indicate the positive and negative arcseconds
  where $0''$ is defined at the target galaxy center.  --- (b) The
  DIS/APO rotation curve of J081420G1 determined from the {\Ha}
  emission line. Below the LRIS/Keck absorption profiles of the
  {\MgIIdblt} doublet which are aligned with the galaxy systemic
  velocity. The quasar continuum fit is indicated by the dashed line
  and the solid line (red) shows the Gaussian fit to the absorption
  features. Below, the solid (green) line shows the sigma
  spectrum. --- (c) Same (a) as except for the J091119.16+031152.9
  quasar field and the physical size of the image at the absorption
  redshift is 178.3 $\times$ 178.3~kpc.  --- (d) Same as (b) except the
  DIS/APO rotation curve is for J091119G1.}
\label{fig:plotVr2}
\end{center}
\end{figure*}

\begin{figure*}
\begin{center}
\includegraphics[angle=90,scale=0.40]{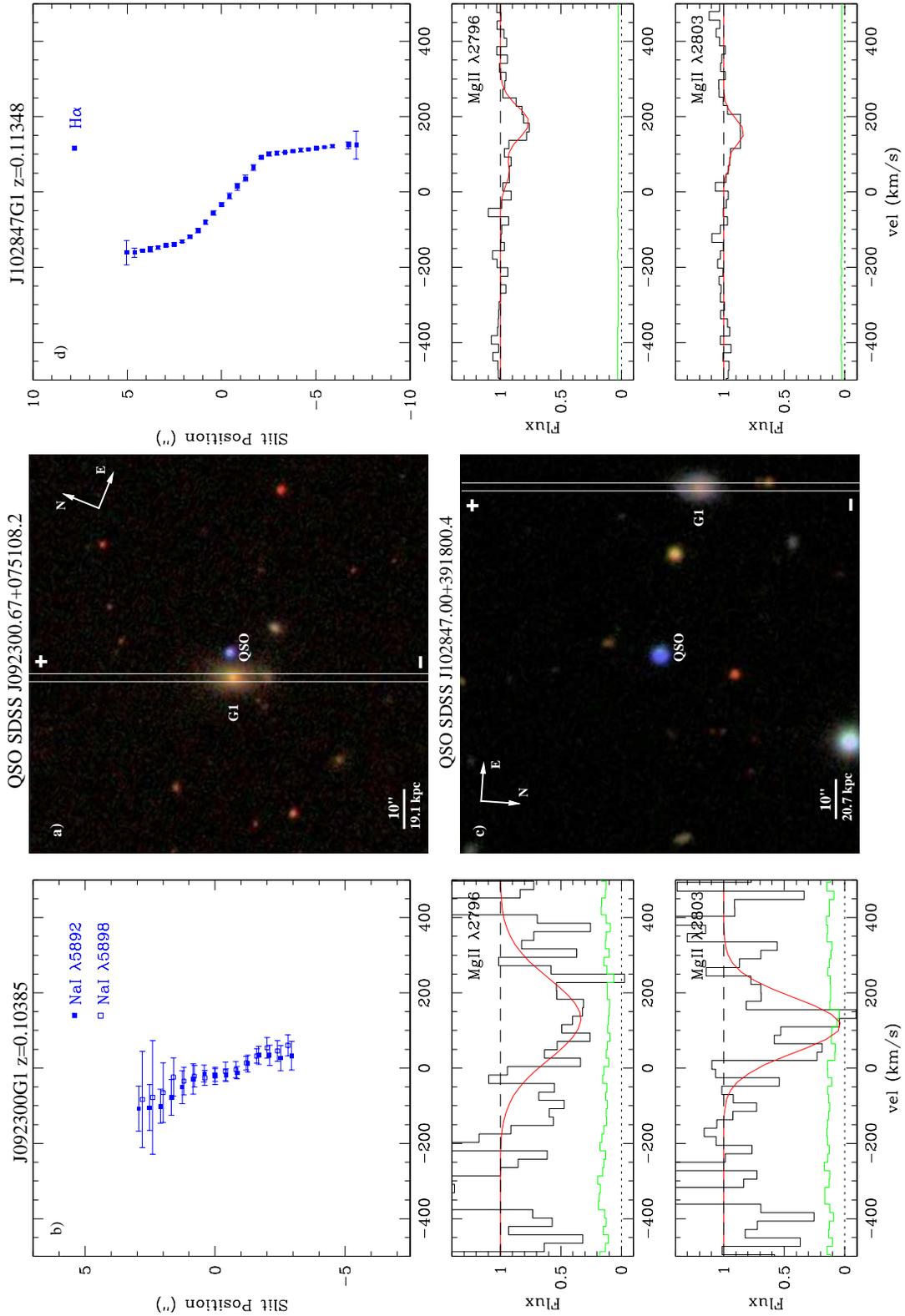}
\caption{(a) A $100'' \times 100''$ (191.4 $\times$ 191.4~kpc at the
  absorption redshift) Sloan $gri$ color image of the SDSS
  J092300.67+075108.2 quasar field. The DIS/APO slit is superimposed on
  the image over the absorbing galaxy J092300G1.  The '$+$' and '$-$'
  on the slit indicate the positive and negative arcseconds where
  $0''$ is defined at the target galaxy center.  --- (b) The DIS/APO
  rotation curve of J092300G1 determined from {\NaID} $\lambda\lambda
  5892$ $5898$ absorption doublet. Below the LRIS/Keck absorption
  profiles of the {\MgIIdblt} doublet which are aligned with the
  galaxy systemic velocity. The quasar continuum fit is indicated by
  the dashed line and the solid line (red) shows the Gaussian fit to
  the absorption features. Below, the solid (green) line shows the
  sigma spectrum. --- (c) Same (a) as except for the
  J102847.00+391800.4 quasar field and the physical size of the image
  at the absorption redshift is 206.8 $\times$ 206.8~kpc.  --- (d)
  Same as (b) except the DIS/APO rotation curve is for J102847G1
  determined from {\Ha}. }
\label{fig:plotVr3}
\end{center}
\end{figure*}

\begin{figure*}
\begin{center}
\includegraphics[angle=90,scale=0.40]{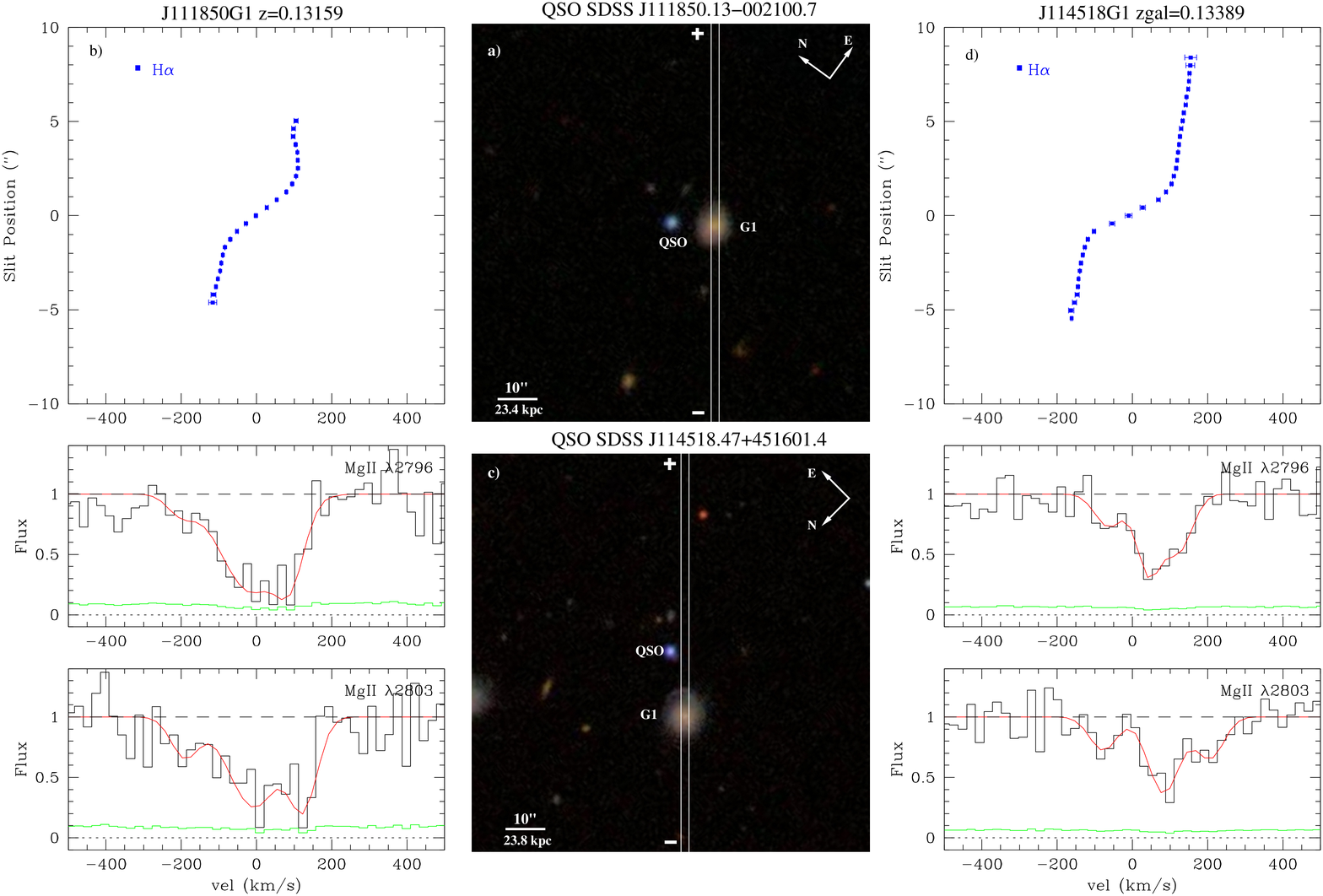}
\caption{(a) A $100'' \times 100''$ (234.2 $\times$ 234.2~kpc at the
  absorption redshift) Sloan $gri$ color image of the SDSS
  J111850.13-002100.7 quasar field. The DIS/APO slit is superimposed
  on the image over the absorbing galaxy J111850G1.  The '$+$' and
  '$-$' on the slit indicate the positive and negative arcseconds
  where $0''$ is defined at the target galaxy center.  --- (b) The
  DIS/APO rotation curve of J111850G1 determined from the {\Ha}
  emission line. Below the LRIS/Keck absorption profiles of the
  {\MgIIdblt} doublet which are aligned with the galaxy systemic
  velocity. The quasar continuum fit is indicated by the dashed line
  and the solid line (red) shows the Gaussian fit to the absorption
  features. Below, the solid (green) line shows the sigma
  spectrum. --- (c) Same (a) as except for the J114518.47+451601.4
  quasar field and the physical size of the image at the absorption
  redshift is 238.0 $\times$ 238.0~kpc. The galaxy directly to the
  left of the absorber, at the edge of the image, is at $z=0.0700$ and
  is not associated with the absorption system at $z_{abs}=
  0.13402$. --- (d) Same as (b) except the DIS/APO rotation curve is
  for J114518G1.}
\label{fig:plotVr4}
\end{center}
\end{figure*}

\begin{figure*}
\begin{center}
\includegraphics[angle=90,scale=0.40]{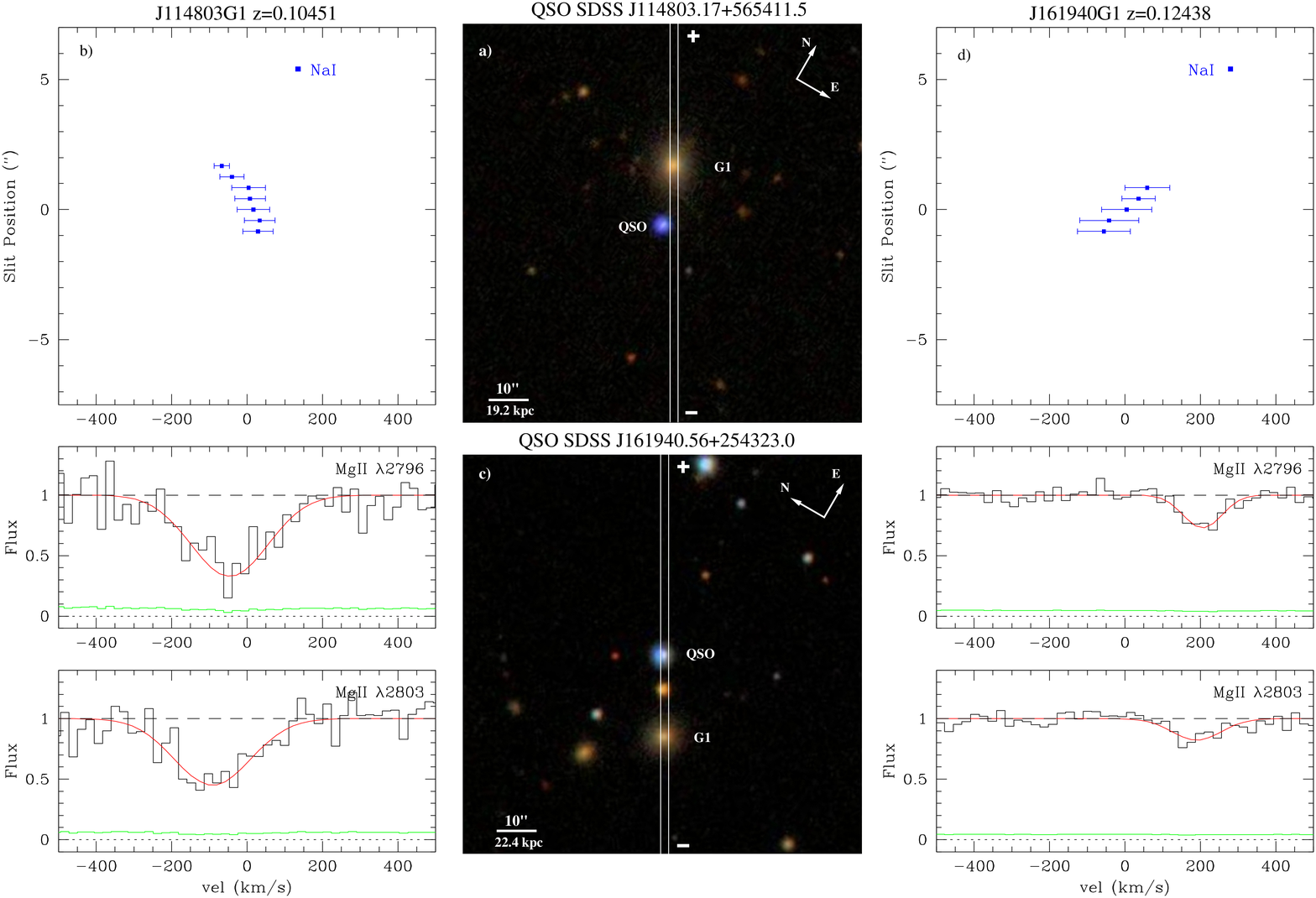}
\caption{(a) A $100'' \times 100''$ (191.6 $\times$ 191.6~kpc at the
  absorption redshift) Sloan $gri$ color image of the SDSS
  J114803.17+565411.5 quasar field. The DIS/APO slit is superimposed
  on the image over the absorbing galaxy J114803G1.  The '$+$' and
  '$-$' on the slit indicate the positive and negative arcseconds
  where $0''$ is defined at the target galaxy center.  --- (b) The
  DIS/APO rotation curve of J114803G1 determined from the {\NaID}
  $\lambda\lambda 5892 5898$~{\AA} absorption doublet, which was fit
  with a single Gaussian.  Below the LRIS/Keck absorption profiles of
  the {\MgIIdblt} doublet which are aligned with the galaxy systemic
  velocity. The quasar continuum fit is indicated by the dashed line
  and the solid line (red) shows the Gaussian fit to the absorption
  features. Below, the solid (green) line shows the sigma
  spectrum. --- (c) Same (a) as except for the J161940.56+254323.0
  quasar field and the physical size of the image at the absorption
  redshift is 224.2 $\times$ 224.2~kpc.  --- (d) Same as (b) except
  the DIS/APO rotation curve determined from the {\Ha} emission line
  is for J161940G1.}
\label{fig:plotVr6}
\end{center}
\end{figure*}

\begin{figure*}
\begin{center}
\includegraphics[scale=0.40]{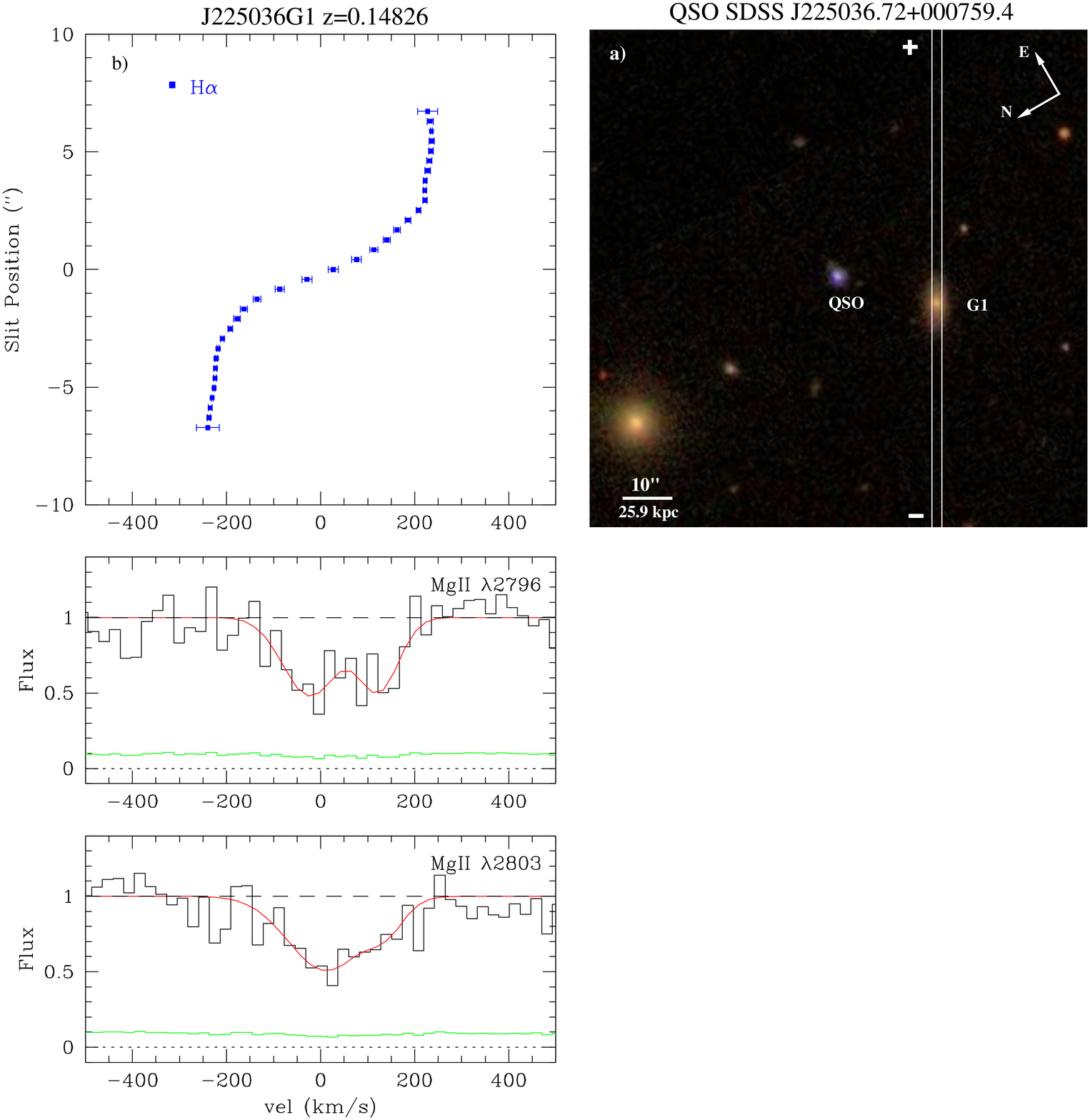}
\caption{(a) A $100'' \times 100''$ (259.2 $\times$ 259.2~kpc at the
  absorption redshift) Sloan $gri$ color image of the SDSS
  J225036.72+000759.4 quasar field. The DIS/APO slit is superimposed
  on the image over the absorbing galaxy J225036G1.  The '$+$' and
  '$-$' on the slit indicate the positive and negative arcseconds
  where $0''$ is defined at the target galaxy center. The galaxy
  directly north of the absorber, in the lower left corner the image,
  is at $z=0.1117$ and is not associated with the absorption system at
  $z_{abs}= 0.14837$. --- (b) The DIS/APO rotation curve of J225036G1
  determined from the {\Ha} emission line. Below the LRIS/Keck
  absorption profiles of the {\MgIIdblt} doublet which are aligned
  with the galaxy systemic velocity. The quasar continuum fit is
  indicated by the dashed line and the solid line (red) shows the
  Gaussian fit to the absorption features. Below, the solid (green)
  line shows the sigma spectrum.}
\label{fig:plotVr7}
\end{center}
\end{figure*}

In Figures~\ref{fig:plotVr2}--\ref{fig:plotVr5} we show combined
$gri-$band SDSS color images of the galaxies and quasar fields.  We
used GIM2D \citep{simard02} to model the galaxy morphologies.  For
each galaxy, the morphological parameters were quantified by fitting a
two-component (bulge+disk) co-spatial parametric model to its
two-dimensional surface brightness distribution. We fit each galaxy
surface brightness profile with a S{\'e}rsic bulge component (with
$0.2 \leq n \leq 4.0$) and an exponential disk component. GIM2D has
been previously used to model $z\sim 0.5$ {\MgII} absorption-selected
galaxies \citep{kacprzak07,kacprzak10c}.  Here, we use GIM2D to
acquire the quasar--galaxy impact parameters $(D)$, inclination angles
($i$), and position angles ($PA$) of the galaxy major axes with
respect to the quasar line-of-sight.

During the GIM2D modeling process, the models are convolved with a
point spread function (PSF) determined by the user. To determine the
2D PSF, we selected ten or more stars close to the absorbing galaxy in
the Sloan $r-$band images and modeled them using DAOPHOT
\citep{stetson87,stetson99}. We then used that PSF in combination with
GIM2D to model SDSS $r-$band galaxy images to extract morphological
properties. We chose to model the galaxies using r-band images since
it has the best sensitivity and it also traces the $\rm{H} \alpha$
emission at $z\sim 0.1$.  The galaxy modeled orientations are found in
Table~\ref{tab:params}.

The magnitudes quoted in Table~\ref{tab:DIS} are k-corrected and
corrected for Galactic reddening \citep{blanton05}. Rest $r-$band
luminosities were computed using $M_r^{\star}=-19.67$ derived for
$z\sim 0.1$ SDSS galaxies \citep{montero-dorta09,blanton03}.



\section{Discussion of Individual Fields}\label{fields}

Here we discuss the halo gas and galaxy kinematics of the 13
absorption-selected galaxies along 11 different quasar sight-lines
shown in Figures~\ref{fig:plotVr2}--\ref{fig:plotVr5}. In
Table~\ref{tab:zdat}, we list all the galaxies in each field that have
spectroscopically confirmed redshifts.  The table columns are (1)
galaxy name, (2) the quasar--galaxy impact parameter, (3) the galaxy
redshift, (4) the {\MgII} absorption redshift, (5) the {\MgII}
absorption and galaxy velocity offset, (6) the rest--frame {\MgII}
$\lambda 2796$ and (7) the {\MgII} $\lambda 2803$ equivalent width,
(8) the doublet ratio, $DR$, and the blue (9) and red (10) velocity
limits of the {\MgII} $\lambda 2796$ absorption profiles.  The galaxy
velocity offsets from the optical-depth-weighted mean {\MgII}
absorption range from $-14$ to $+212$~{\kms}.  Galaxy redshifts will
only be quoted to four significant figures from here on for
simplicity.  We will later discuss kinematic halo models in
\S~\ref{halo}.

\subsection{{\MgII} Absorption from Isolated Galaxies}

In the following sub-sections we discuss the nine galaxy-absorber
pairs which appear to be isolated systems and do not have any major or
minor companions within 100~kpc of the quasar line-of-sight. We
discuss the remaining four absorbers in Section~\ref{sec:double}.  In
all Figures~\ref{fig:plotVr2}--\ref{fig:plotVr5}, we show a $100''
\times 100''$ SDSS image of the field centered on the quasar.  Below
the rotation curve of each galaxy, the {\MgIIdblt} absorption profiles
are shown on the same velocity scale. For each galaxy, the slit
position angle was selected to lie along the galaxy major axis.

\newpage

\subsubsection{J081420G1}

In Figure~\ref{fig:plotVr2}a we show that the absorption system
detected in the spectrum of the background quasar is associated with a
$1.51L^{\star}_r$ galaxy located 51~kpc away from the quasar
line-of-sight. This spiral galaxy is moderately inclined at 40
degrees.

The rotation curve of G1, presented in Figure~\ref{fig:plotVr2}b, is
derived from {\Ha} and exhibits a maximum projected rotation velocity
of $\sim$130~{\kms}.  The mean absorption redshift is offset by
$+$105~{\kms} from the galaxy systemic velocity. The {\MgII}
absorption is an apparent single kinematic component having a velocity
spread of roughly 200~\kms.  The doublet ratio suggests that this
$W_r(2796)=0.57$~{\AA} system is not saturated. The {\MgII}
absorption resides to one side of the galaxy systemic velocity and
also aligns with the maximum rotation velocity of the galaxy.

\subsubsection{J091119G1}

In Figure~\ref{fig:plotVr2}c we show that the absorption system
detected in the spectrum of the background quasar is associated with
an almost edge-on ($i=$82$^\circ$) $1.38L^{\star}_r$ spiral galaxy
located 72~kpc from the quasar line-of-sight.

The rotation curve of G1, presented in Figure~\ref{fig:plotVr2}d, is
derived from {\Ha} and exhibits a maximum projected rotation velocity
of $\sim$230~{\kms}. The mean absorption redshift is offset by
$+$66~{\kms} from the galaxy systemic velocity. The {\MgII} absorption
consists of two broad regions that span roughly 790~\kms.  The doublet
ratio suggests that that this $W_r(2796)=0.82$~{\AA} system is not
saturated. The galaxy velocities are consistent with most of the
absorbing gas velocities with the two highest optical depth regions
aligning with both sides of the galaxy rotation curve. 

\subsubsection{J092300G1}

In Figure~\ref{fig:plotVr3}a we show that the absorption system
detected in the spectrum of the background quasar is associated with
an $2.27L^{\star}_r$ S0-like galaxy located only 12~kpc from the
quasar line-of-sight. The galaxy has an inclination of 56 degrees.

The dynamics of G1, presented in Figure~\ref{fig:plotVr3}b, is derived
from {\NaID} absorption and exhibits a maximum projected rotation
velocity of $\sim$110~{\kms}.  The mean absorption redshift is offset
by $+$127~{\kms} from the galaxy systemic velocity. The {\MgII}
absorption comprises a single component that span roughly 605~\kms,
with the bulk of the gas spanning $\sim$200~{\kms}. The doublet ratio
suggests that this strong $W_r(2796)=2.25$~{\AA} systems is partially
saturated. The bulk of the absorption resides to one side of the
galaxy rotation curve, although it tends to have slightly higher
velocities than the maximum rotation.

\subsubsection{J102847G1}

In Figure~\ref{fig:plotVr3}c we show that the absorption system
detected in the spectrum of the background quasar is associated with
an $1.49L^{\star}_r$ spiral galaxy located 90~kpc from the quasar
line-of-sight. The galaxy has an inclination of 54 degrees.

The rotation curve of G1, presented in Figure~\ref{fig:plotVr3}d, is
derived from {\Ha} and exhibits a maximum projected rotation velocity
of $\sim$160~{\kms}.  The mean absorption redshift is offset by
$+$168~{\kms} from the galaxy systemic velocity. The {\MgII}
absorption consists of a broad shallow and single narrower component
that span roughly 385~\kms. The doublet ratio suggests that this weak
$W_r(2796)=0.30$~{\AA} system is saturated. The bulk of the
absorption resides to one side of the galaxy systemic velocity, and
aligns with the maximum rotation of the galaxy.

\subsubsection{J111850G1}

In Figure~\ref{fig:plotVr4}a we show that the absorption system
detected in the spectrum of the background quasar is associated with
an $1.91L^{\star}_r$ spiral galaxy located only 25~kpc from the quasar
line-of-sight. The galaxy has an inclination of 30 degrees.

The rotation curve of G1, presented in Figure~\ref{fig:plotVr4}b, is
derived from {\Ha} and exhibits a maximum projected rotation velocity
of $\sim$120~{\kms}. The mean absorption redshift is offset by
$-$5~{\kms} from the galaxy systemic velocity. The {\MgII} absorption
consists of a large broad component that spans roughly 480~\kms. The
doublet ratio suggests that this strong $W_r(2796)=1.93$~{\AA} system
is mostly saturated. The bulk of the absorption spans the entire
velocity range of the galaxy rotation curve almost centered on the
galaxy systemic velocity.

\subsubsection{J114518G1}
In Figure~\ref{fig:plotVr4}c we show that the absorption system
detected in the spectrum of the background quasar is associated with
an $2.23L^{\star}_r$ spiral galaxy located 34~kpc from the quasar
line-of-sight. The galaxy has an inclination of 34 degrees.

The rotation curve of G1, presented in Figure~\ref{fig:plotVr4}d, is
derived from {\Ha} and exhibits a maximum projected rotation velocity
of $\sim$160~{\kms}.  The mean absorption redshift is offset by
$+$46~{\kms} from the galaxy systemic velocity. The {\MgII} absorption
consists of a large broad component that spans roughly 437~\kms. The
doublet ratio suggests that this strong $W_r(2796)=1.06$~{\AA} system
is mostly saturated. The bulk of the absorption resides primarily to
one side of the galaxy systemic velocity and spans from the galaxy
systemic velocity to the maximum rotation velocity of the galaxy.

\subsubsection{J114803G1}

In Figure~\ref{fig:plotVr6}a we show that the absorption system
detected in the spectrum of the background quasar is associated with
an $2.17L^{\star}_r$ elliptical galaxy located 29~kpc from the quasar
line-of-sight.

The spatial radial velocity of G1, as derived from {\NaI}, is
presented in Figure~\ref{fig:plotVr6}b. The data suggest that G1
exhibits more of a global shear than rotation and has a maximum
observed shear velocity of $\sim$67~{\kms}.  The mean absorption
redshift is offset by $-$62~{\kms} from the galaxy systemic velocity.
The {\MgII} absorption consists of a large broad component that spans
roughly 558~\kms. The doublet ratio suggests that this strong
$W_r(2796)=1.59$~{\AA} system is mostly saturated. The bulk of the
absorption spans the entire velocity range of the galaxy rotation
curve almost centered on the galaxy systemic velocity.

\subsubsection{J161940G1}

In Figure~\ref{fig:plotVr6}c we show that the absorption system
detected in the spectrum of the background quasar is associated with
an $1.70L^{\star}_r$ elliptical galaxy located 46~kpc from the quasar
line-of-sight.

The spatial radial velocity of G1, as derived from {\NaID}, is
presented in Figure~\ref{fig:plotVr6}d. The data suggest that G1
exhibits more of a global shear than rotation and has a maximum
observed shear velocity of $\sim$70~{\kms}. The mean absorption
redshift is offset by $+$212~{\kms} from the galaxy systemic velocity.
The {\MgII} absorption consists of a single narrow component that
spans roughly 298~\kms. The doublet ratio suggests that this weaker
$W_r(2796)=0.32$~{\AA} system is mostly saturated. The absorption
resides to one side of the galaxy systemic velocity.

\subsubsection{J225036G1}

In Figure~\ref{fig:plotVr7}a we show that the absorption system
detected in the spectrum of the background quasar is associated with
an $4.27L^{\star}_r$ elliptical galaxy located 54~kpc from the quasar
line-of-sight.

The rotation curve of G1, presented in Figure~\ref{fig:plotVr7}b, is
derived from {\Ha} and exhibits a maximum projected rotation velocity
of $\sim$240~{\kms}.  The mean absorption redshift is offset by
$+$38~{\kms} from the galaxy systemic velocity. The {\MgII} absorption
consists of a single narrow component that spans roughly 365~\kms. The
doublet ratio suggests that this strong $W_r(2796)=1.08$~{\AA} system
is mostly saturated. The absorbing gas spans both sides of the galaxy
systemic velocity, however the bulk of the absorption resides to one
side.

\begin{figure*}
\begin{center}
\includegraphics[angle=90,scale=0.40]{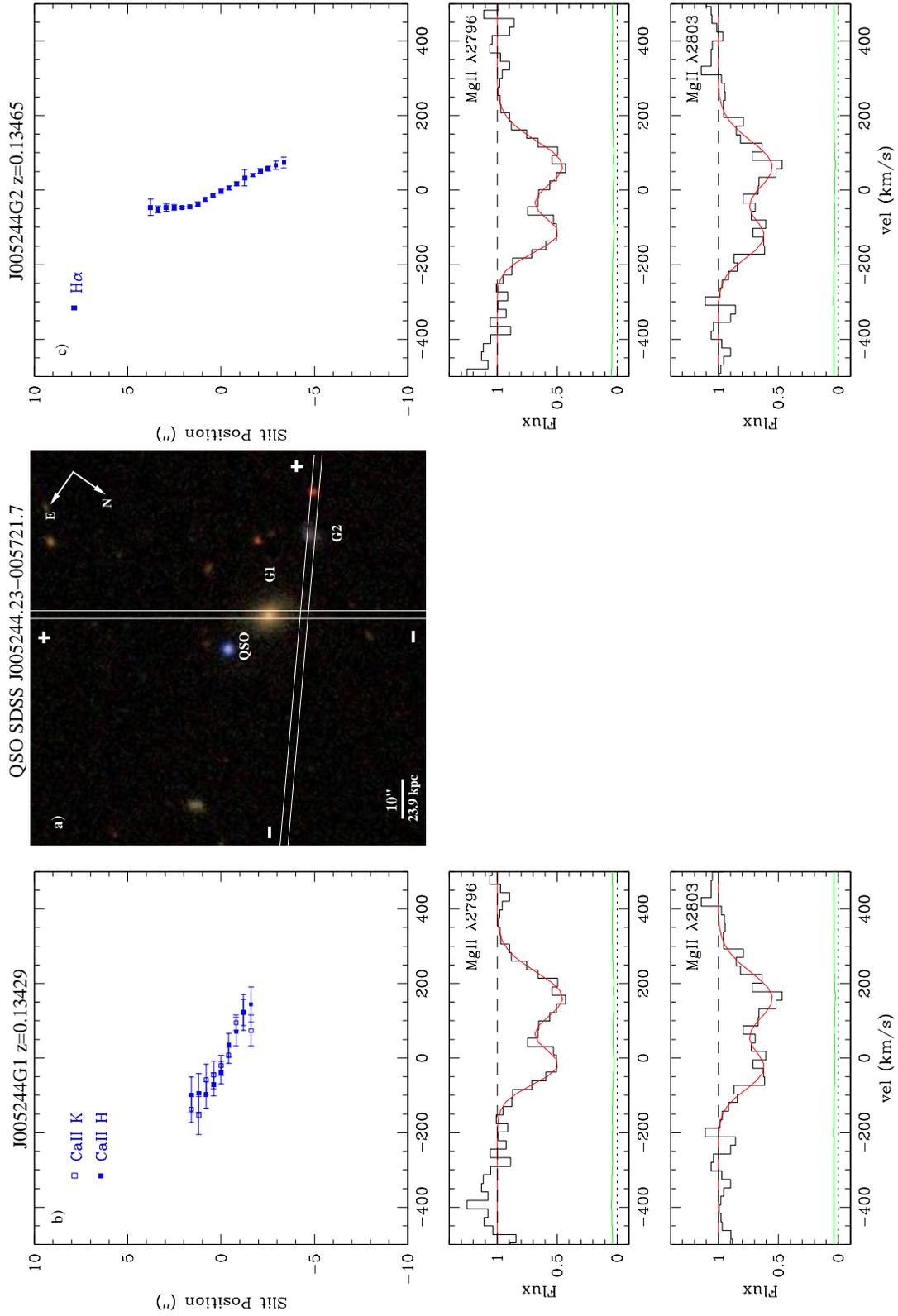}
\caption{(a) A $100'' \times 100''$ (238.8 $\times$ 238.8~kpc at the
  absorption redshift) Sloan gri color image of the quasar field. The
  DIS/APO slit is superimposed on the image.  The '$+$' and '$-$' on
  the slit indicate the positive and negative arcseconds where $0''$
  is defined at the target galaxy center.  This field contains two
  absorbing galaxies labeled as G1 and G2. An additional galaxy
  located at the far South East of the quasar is not part of this
  group and has a redshift of $z=0.1390\pm 0.0007$ which is more than
  $1250$~{\kms} redward of the galaxy pair. --- (b) The DIS/APO
  rotation curve of G1 and the LRIS/Keck absorption profiles aligned
  with the galaxy systemic velocity. The zeropoint velocity is set by
  G1. The quasar continuum fit is indicated by the dashed line and the
  solid line (red) shows the Gaussian fit to the absorption
  features. Below, the solid (green) line shows the sigma
  spectrum. (c) Same as (b) except the DIS/APO rotation curve
  determined from the {\Ha} emission line is for G2.}
\label{fig:plotVr1}
\end{center}
\end{figure*}

\begin{figure*}
\begin{center}
\includegraphics[angle=90,scale=0.40]{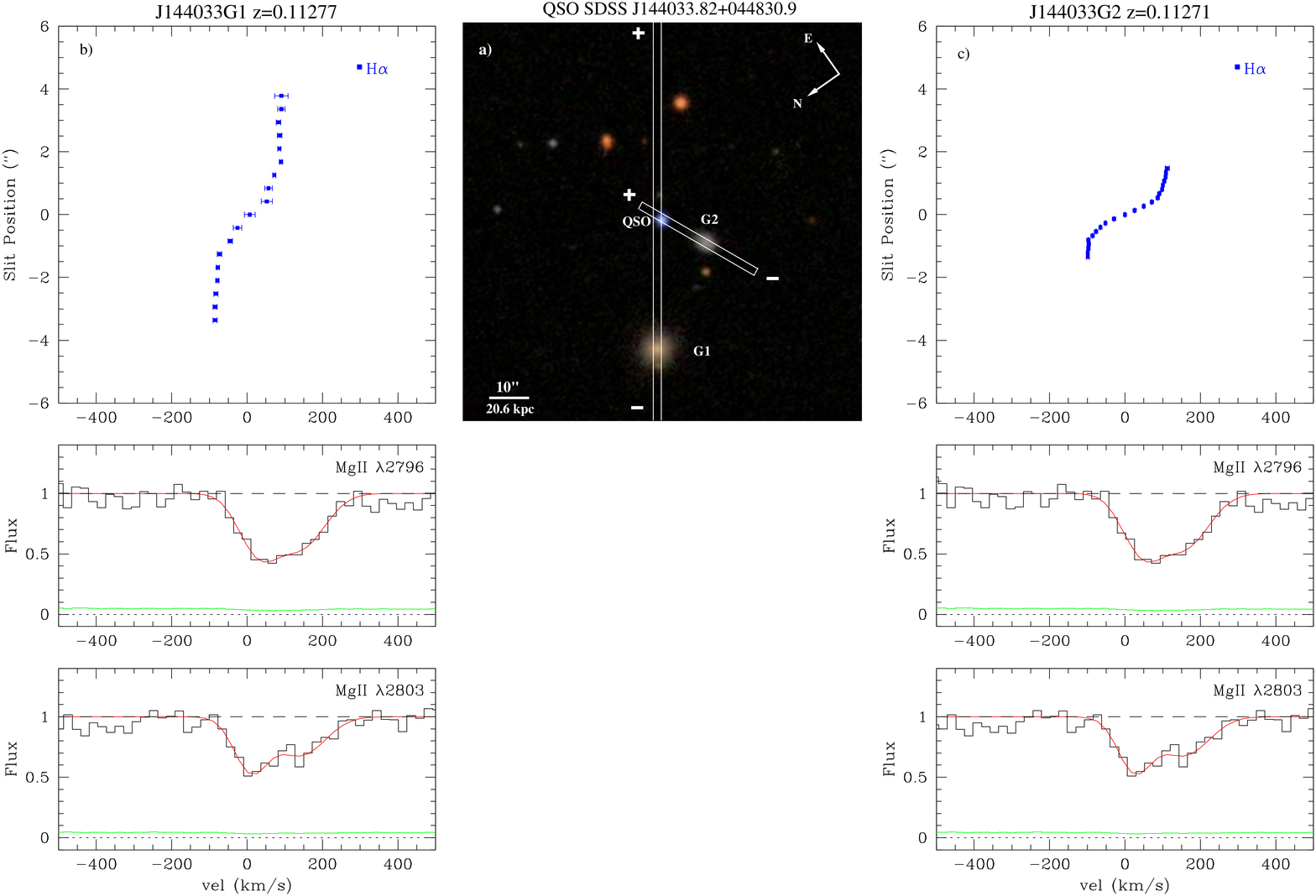}
\caption{(a) A $100'' \times 100''$ (205.6 $\times$ 205.6~kpc at the
  absorption redshift) Sloan gri color image of the quasar field. The
  DIS/APO slit is superimposed on the image for G1 and the ESI/Keck
  $20''$ slit superimposed over G2.  The '$+$' and '$-$' on the slit
  indicate the positive and negative arcseconds where $0''$ is defined
  at the target galaxy center.  This field contains two absorbing
  galaxies labeled as G1 and G2.  --- (b) The DIS/APO rotation curve
  of G1 and the LRIS/Keck absorption profiles aligned with the galaxy
  systemic velocity. The zeropoint velocity is set by G1. The quasar
  continuum fit is indicated by the dashed line and the solid line
  (red) shows the Gaussian fit to the absorption features. Below, the
  solid (green) line shows the sigma spectrum.  (c) Same as (b) except
  the rotation curve for G2 is obtained with ESI/Keck.}
\label{fig:plotVr5}
\end{center}
\end{figure*}

\subsection{{\MgII} Absorption from Galaxy Pairs/Groups}\label{sec:double}

Here we discuss galaxies in ``group'' environments which are typically
defined by two or more galaxies within the standard halo size of
$\sim$120~kpc \citep{chen08,kacprzak08,chen10a}. It has been suggested
that group environments may give rise to {\MgII} absorption from a
range of structures such as from tidal tails, streams, etc.
\citep{bowen95,cwc99,whiting06,kacprzak10a,kacprzak10b}.  It has also
been suggested that galaxy pairs/groups may not follow the same clear
anti-correlation between $D$ and $W_r(2796)$ as seen for isolated
galaxies \citep{chen10a}. Here we discuss the two absorbing pairs of
galaxies.

\subsubsection{J005244G1 \& J005244G2}

In Figure~\ref{fig:plotVr1}a we show that the absorption system
detected in the spectrum of the background quasar maybe associated with
two galaxies. G1 is a $2.86L^{\star}_r$ elliptical galaxy located
32~kpc from the quasar line-of-sight and G2 is a $0.23L^{\star}_r$
emission-line galaxy located 86~kpc from the quasar line-of-sight.  In
the short exposure SDSS image, there is no evidence of galaxy--galaxy
interaction.

The rotation curve of G2, presented in Figure~\ref{fig:plotVr1}b, is
derived from {\CaII} absorption and exhibits a maximum projected
rotation velocity of $\sim$143~{\kms}.  The mean absorption redshift
is offset by $-$15~{\kms} from the galaxy systemic velocity.  The
rotation curve of G2, presented in Figure~\ref{fig:plotVr1}c, is
derived from {\Ha} and exhibits a maximum projected rotation velocity
of $\sim$74~{\kms}.  Below the {\MgIIdblt} absorption profiles are
shown with the systemic velocity of G1 is the velocity zeropoint.  The
mean absorption redshift is offset by $+$82~{\kms} from the galaxy
systemic velocity. We find that the absorption spans both sides of the
systemic velocity for both galaxies.

The {\MgII} absorption has a broad profile with two clear kinematic
components that spans roughly 596~{\kms}. The doublet ratio suggests
that this strong $W_r(2796)=1.43$~{\AA} system is partially
saturated.

\subsubsection{J144033G1 \& J144033G2}

In Figure~\ref{fig:plotVr5}a we show that the absorption system
detected in the spectrum of the background quasar maybe associated
with two spiral galaxies: G1 is a $1.72L^{\star}_r$ galaxy located
67~kpc from the quasar line-of-sight and G2 is a $0.56L^{\star}_r$
galaxy located 25~kpc from the quasar line-of-sight.  In the short
exposure SDSS image, there is no clear evidence of galaxy--galaxy
interaction.

The rotation curve of G1, presented in Figure~\ref{fig:plotVr5}b, is
derived from {\Ha} and exhibits a maximum projected rotation velocity
of $\sim$91~{\kms}. The mean absorption redshift is offset by
$+$91~{\kms} from the galaxy systemic velocity.

The rotation curve of G2, presented in Figure~\ref{fig:plotVr5}c is
also derived from {\Ha} and exhibits a maximum projected rotation
velocity of $\sim$112~{\kms}.  Below the {\MgIIdblt} absorption
profiles are shown with the systemic velocity of G1 is the velocity
zeropoint.  The mean absorption redshift is offset by $+$89~{\kms}
from the galaxy systemic velocity.  The absorption resides to one side
of systemic velocity of both galaxies.

The {\MgII} absorption consists of a broad single kinematic component
that spans roughly 408~{\kms}. The doublet ratio suggests that this
$W_r(2796)=1.18$~{\AA} system is mostly saturated.

\subsection{Summary I: Observational Kinematic Comparisons}

We find that for the nine isolated galaxy--absorber pairs, seven
galaxies have well defined rotation curves while two galaxies display
only shear. We find that in only 3/9 cases the absorption resides to
one side of the galaxy systemic velocity and the absorption redshift
tends to align with one side of the rotation curve. In the remaining
6/9 cases, the absorption spans both sides of the galaxy systemic
velocity, although the bulk of the {\MgII} resides mostly to one side
of the galaxy systemic velocity.

For our double galaxy--absorber pairs, we find that all four galaxies
exhibit well defined rotation curves. In one case, the absorbing gas
spans both sides of both host galaxy systemic velocities. In the other
case, the absorbing gas resides to one side of the systemic velocity
of both absorbing galaxies.

In the next section we explore if extended disk-like halo rotation can
explain the distribution of absorption kinematics detected at a range
of impact parameters from the host galaxies.

\begin{deluxetable}{llllrc}
\tabletypesize{\scriptsize}
\tablecaption{Galaxy Model Parameters and Disk Model Input Values\label{tab:params}}
\tablecolumns{6}
\tablewidth{0pt}

\tablehead{
\colhead{Galaxy}&
\colhead{$D$} &
\colhead{$i$} &
\colhead{$PA$} &
\colhead{$v_{max}$} &
\colhead{$r_h$} \\
\colhead{Name}&
\colhead{(kpc)} &
\colhead{(deg.)}&
\colhead{(deg.)} &
\colhead{(km/s)} &
\colhead{(kpc)}
}
\startdata
J005244G1   & $32.4\pm0.2$ & $45_{- 3}^{+ 5}$ &  $59_{- 6}^{+ 7}$ & 143.9& 5.91 \\
            &                   &       &                 &              &      \\[-1.0ex]
J005244G2 & $86.1\pm1.2$ & $42_{-10}^{+10}$ &  $65_{-31}^{+25}$ & $-$73.5& 3.57 \\
            &                   &       &                 &              &      \\[-1.0ex]
J081420G1 & $51.1\pm0.3$ & $40_{- 2}^{+ 1}$ &  $68_{- 2}^{+ 3}$ & 131.4  & 7.18 \\
            &                   &       &                 &              &      \\[-1.0ex]
J091119G1 & $72.1\pm0.4$ & $82_{- 1}^{+ 1}$ &  $ 6_{- 1}^{+ 0}$ & 231.6  & 7.84 \\
            &                   &       &                 &              &      \\[-1.0ex]
J092300G1 & $11.9\pm0.3$ & $56_{- 2}^{+ 2}$ &  $25_{- 3}^{+ 3}$ & $-$107.7& 5.96 \\
            &                   &       &                 &              &      \\[-1.0ex]
J102847G1 & $89.8\pm0.4$ & $54_{- 2}^{+ 2}$ &  $62_{- 1}^{+ 2}$ & 161.6  & 6.44 \\
            &                   &       &                 &              &      \\[-1.0ex]
J111850G1 & $25.1\pm0.3$ & $30_{- 2}^{+ 2}$ &  $29_{- 2}^{+ 0}$ & $-$116.2& 6.31 \\
            &                   &       &                 &              &      \\[-1.0ex]
J114518G1 & $39.4\pm0.8$ & $34_{- 2}^{+ 2}$ &  $74_{- 5}^{+ 4}$ & $-$161.5& 6.90 \\
            &                   &       &                 &              &      \\[-1.0ex]
J114803G1 & $29.1\pm0.5$ & $45_{- 3}^{+ 4}$ &  $62_{- 4}^{+ 4}$ & $-$67.2& 5.63 \\
            &                   &       &                 &              &      \\[-1.0ex]
J144033G1 & $67.1\pm0.1$ & $30_{- 3}^{+ 3}$ &  $89_{- 4}^{+ 6}$ & $-$90.7& 4.62 \\
            &                   &       &                 &              &      \\[-1.0ex]
J144033G2 & $24.9\pm0.2$ & $55_{- 5}^{+ 3}$ &  $38_{- 8}^{+ 7}$ & $-$111.8& 2.99 \\
            &                   &       &                 &              &      \\[-1.0ex]
J161940G1 & $45.7\pm0.7$ & $12_{-12}^{+12}$ &  $45_{-28}^{+47}$ & 59.1   & 4.49 \\
            &                   &       &                 &              &      \\[-1.0ex]
J225036G1 & $53.9\pm0.7$ & $70_{- 2}^{+ 1}$ &  $15_{- 1}^{+ 1}$ & $-$239.7&4.96  \\
            &                   &       &                 &         
\enddata
\end{deluxetable}

\section{Galaxy Kinematics and Halo--Disk Models}\label{halo}

We now apply the simple monolithic halo model of \citet{steidel02} to
determine whether an extended disk-like rotating halo is able to
account for all or most of the observed {\MgII} absorption velocity
spread measured in our galaxy/absorber systems (see Steidel et
al. 2002 for a detailed description of the model). Here we briefly
describe the model, which treats the halo gas as a co-rotating thick
disk with decreasing velocity as a function of scale height.

The line-of-sight disk-halo velocity, $v_{los}$, is dependent on four
measurable quantities, $D$, $i$, $PA$, and $v_{max}\equiv\mbox{maximum
projected galaxy rotation velocity}$, such that

\begin{eqnarray} 
\label{eq:kine}
v_{los}(y)&=&\frac{-v_{max}}{\sqrt{1+\left(\displaystyle \frac{y}{p}\right)^2}}\mbox{~}
\exp \left\{-\frac{\left|y-y_{\circ}\right|}{h_{v}\tan i}\right\}\mbox{~~}\mbox{where}, \\
\nonumber\\[2.0ex]
y_{\circ}&=&\frac{D\sin PA}{\cos i} \mbox{~~~~~}\mbox{and}\mbox{~~~~~}p=D\cos PA\mbox{~},\nonumber
\end{eqnarray}

\noindent where the free parameter, $h_{v}$, is the lagging halo gas
velocity scale height. The line-of-sight velocity is a function of
$y$, which is the projected line-of-sight position above the
disk-plane, and the parameter $y_{\circ}$ represents the position at
the projected mid-plane of the disk. The range of $y$ is constrained
by the model halo thickness, $H_{\rm eff}$, such that
$y_{\circ}-H_{\rm eff}\tan i \leq y \leq y_{\circ}+H_{\rm eff}\tan i$.
The distance along the line-of-sight relative to the point were it
intersects the projection of the disk mid--plane is then
$D_{los}=(y-y_{\circ})/\sin i$, thus, $D_{los}\equiv 0$ at the disk
mid plane. There are no assumptions about the spatial density
distribution of {\MgII} absorbing gas, except that $H_{\rm eff}$ is
the effective gas layer thickness capable of giving rise to
absorption.

Here we set $h_v=1000$~kpc in order to {\it maximize} the rotational
velocity predicted by the model, which effectively removes the lagging
halo velocity component (such that the exponential in
Equation~\ref{eq:kine} is roughly equal to unity).

\begin{figure}
\begin{center}
\includegraphics[scale=0.90]{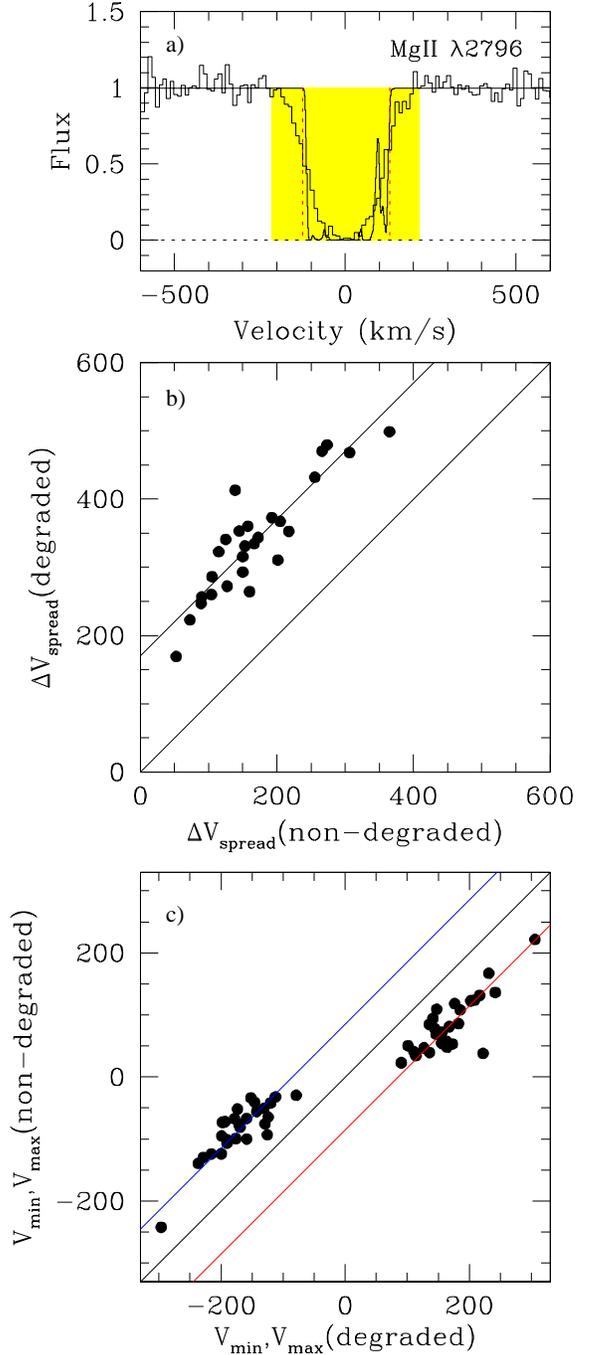}
\caption{(a) The solid line shows an example of a Voigt profile fit to
a HIRES/Keck {\MgII} $\lambda 2796$ absorption system obtained from
\citet{kacprzak10c}. The histogram data shows the high resolution
Voigt profile fit degraded to the resolution typical LRIS/Keck
data. The vertical dashed (red) lines indicate the velocity width of
the {\MgII} $\lambda 2796$ as measured from the HIRES/Keck data and
the highlighted region shows the velocity width of the degraded
spectrum. --- (b) The velocity widths of the 26 high-resolution
{\MgII} $\lambda 2796$ absorption profiles obtained from
\citet{kacprzak10b,kacprzak10c}, as measured from HIRES and UVES
spectra (non-degraded), versus the velocity widths measured from the
degraded synthetic LRIS spectra (degraded). We find that the
resolution, pixelization, and noise in the observed spectra introduce
an average apparent velocity spread increase of $\sim$170~{\kms} with
a scatter of $\sim$60~{\kms}.  --- (c) The shift in velocity due the
degrading of the spectra from both the blue (min) and red (max)
velocity edges (or wings) of the absorption profiles. We find a
symmetric shift of 85~{\kms} with a scatter of 25~\kms.}
\label{fig:mock}
\end{center}
\end{figure}

\subsection{Dealing With Spectral Resolution}

In order to compare our present low-redshift results directly to the
kinematic studies of \citet{kacprzak10a} and \citet{steidel02}
performed at intermediate redshift, we must account for the difference
in the spectral resolution of the two samples. Here, our LRIS spectra
have a velocity resolution of $v \simeq 155~{\kms}$, whereas the
\citet{kacprzak10a} HIRES/Keck and UVES/VLT spectra have a velocity
resolution of $v \simeq 6~{\kms}$.

To compute the velocity broadening of the LRIS {\MgII} absorption
profiles due to spectral resolution, we degraded 26 HIRES/Keck and
UVES/VLT $W_r(2796) > 0.2$~{\AA} systems from \citet{kacprzak10c} and
\citet{kacprzak10b} to have identical resolution of the LRIS data
[these include the systems of the kinematics studies of
\citet{kacprzak10a} and \citet{steidel02}].  We utilized the Voigt
profile parameters (column densities, $b$ parameters, and velocities)
from the Voigt profile fits to the Kacprzak {\etal} spectra and
generated synthetic LRIS spectra of the profiles.  We convolved the
smooth Voigt profile model with a Gaussian instrumental spread of LRIS
($R=3500$) and sampled the spectra with a pixel size of 0.18~{\AA}.
We introduced noise using a signal--to--noise ratio 15 per pixel
(using Gaussian deviates). An uncertainty spectrum, $\sigma
(\lambda)$, is generated that accounts for the LRIS read noise (which
affects the noise in the line cores), using a Poisson plus read-noise,
model \citep[see][]{cwcthesis}
\begin{equation}
\sigma (\lambda) = I_c^{-1} 
\left[ F(\lambda)I_c + {\rm RN}^2 \right] ^{1/2} , 
\end{equation} 
where $F(\lambda)$ is relative counts in the synthesized spectra at
wavelength $\lambda$ and the continuum $I_c$ is
\begin{equation}
I_c = \frac{{\rm SNR}^2}{2} \left\{ 
1 + \left[ 1 + \left( \frac{2{\rm RN}}{{\rm SNR}} \right) ^2 \right] ^{1/2} 
\right\} ,
\end{equation}
where SNR is the signal-to-noise ratio and ${\rm RN}$ is the read
noise.

We generated synthetic LRIS spectra of both members of the {\MgII}
doublet and then fully analyze these spectra in an identical fashion
performed for the LRIS data used in this work.  We thus obtained the
equivalent widths, double ratios, velocity moments, and velocity
spreads (and all uncertainties) for these synthetic LRIS spectra. In
Figure~\ref{fig:mock}a we show an example of the degraded spectra,
noting the symmetric broadening of the absorption profile.

In Figure~\ref{fig:mock}b, we show the distribution of absorption
velocity widths for the degraded and non-degraded systems.  We find
that, using the identical measurement standards for both the synthetic
LRIS spectra and the observed LRIS spectra, the resolution,
pixelization, and noise in the observed spectra introduce an average
apparent velocity spread increase of $\sim$170~{\kms} with a scatter
of $\sim$60~{\kms}.  Furthermore, Figure~\ref{fig:mock}c shows that
the average apparent increase of velocity is symmetric for both the
blue and red wings (edges) of the absorption profiles.  We find an
average velocity increase of $\pm 85$~{\kms} (to the red and to the
blue) relative to the Voigt profile models of the absorbers observed
at HIRES and UVES resolution.  There is a scatter of approximately
25~{\kms} for the 26 systems we examined from the Kacprzak {\etal}
spectra.

We have now applied a resolution correction such that the LRIS {\MgII}
absorption velocity widths translated to observed HIRES or UVES
velocity widths.

\subsection{Results of the Halo--Disk Models}

\begin{figure*}
\begin{center}
\includegraphics[angle=90,scale=0.80]{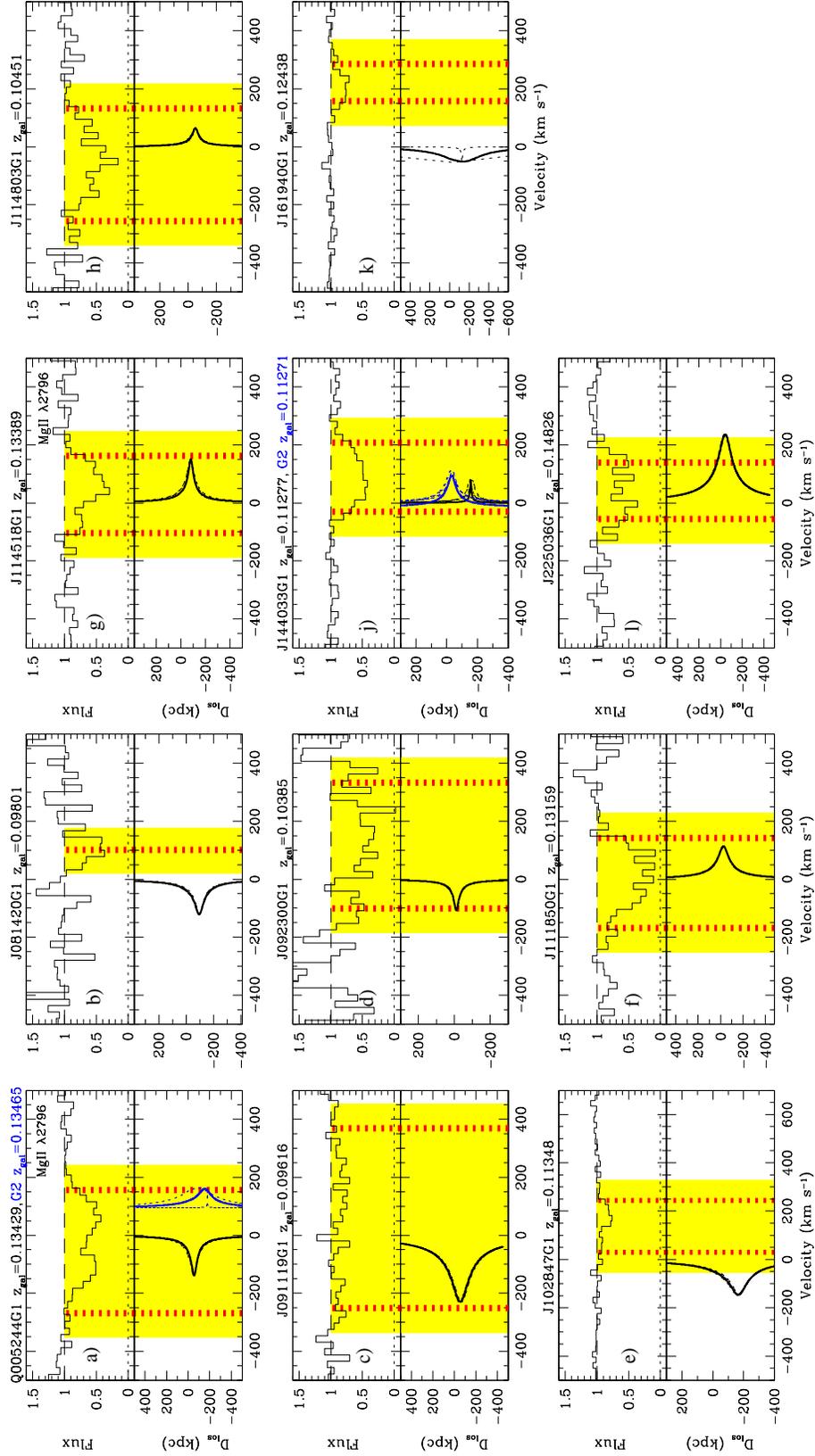}
\caption{The {\MgII} absorption profiles and the disk model velocities
as a function of $D_{los}$ (solid curve) are shown for each galaxy in
the top and bottom panels, respectively. The solid curve is computed
using Equation~\ref{eq:kine} and the values from
Table~\ref{tab:params}. The dashed curves are models computed for the
maximum and minimum predicted model velocities given the uncertainties
of $i$ and $PA$. The {\MgII} absorption velocities are shaded in.  The
thick disk halo model is successful at predicting the observed {\MgII}
absorption velocity distribution when the solid (or dashed) curves
span the same velocity spread as that of the {\MgII} absorption gas
defined by shaded region between the vertical dashed (red) lines. The
vertical dashed (red) lines are an applied $85$~{\kms} resolution
correction such that the LRIS {\MgII} absorption velocity widths
translated to observed HIRES or UVES velocity widths. The thickness of
the vertical dashed (red) lines indicates the $\pm25$~{\kms}
uncertainty in the velocity correction.  The $D_{los}$ is equal to
zero when the quasar line of sight intersects the projected mid--plane
of the galaxy. The panels are as follows; (a) J005244G1 and G2 (b)
J081420G1, (c) J091119G1 (d) J092300G1, (e) J102847G1, (f) J111850G1,
(g) J114518G1, (h) J114803G1, (i) J144033G1 and G2, (k) J161940G1, and
(l) J225036G1.}
\label{fig:toym}
\end{center}
\end{figure*}

In Figure~\ref{fig:toym}$a-l$, we show the {\MgII} absorption profiles
for each galaxy, where the shaded regions indicate detected
absorption.  We have applied a resolution correction such that the
{\MgII} velocity widths (shaded region) are translated to ``observed''
HIRES or UVES velocity widths as indicated by the vertical dashed
lines (red).

Below each absorption profile is the thick disk halo model velocities
as a function of $D_{los}$ derived for each galaxy (solid line) using
Equation~\ref{eq:kine} and parameters in
Table~\ref{tab:params}. Recall that, at $D_{los}=0$~kpc, the model
line of sight intersects the projected mid-plane of the galaxy.  The
dashed curves represent the range of disk halo model velocities
derived from the combination of the minimum and maximum uncertainties
in the $PA$ and $i$. In some cases the values of the $PA$ and $i$ are
well determined such that the dashed curves lie on the solid curves
(see Figure~\ref{fig:toym}$b$).  The model also predicts the
line-of-sight position, $D_{los}$, of the halo gas at each velocity,
$v_{los}$.

The thick disk halo model is successful at predicting the observed
{\MgII} absorption velocity distribution when the solid (or dashed)
curves span the same velocity spread as that of the {\MgII} absorption
gas defined by shaded region between the vertical dashed (red) lines.
If this is not the case, one can conclude that disk-like halo rotation
is not the {\it only\/} dynamic mechanism responsible for the {\MgII}
kinematics. In the following subsections we discuss the halo models of
the individual galaxies.

\subsubsection{J005244G1 \& J005244G2}

Galaxies G1 and G2 are potentially interacting galaxies, given their
close angular proximity and redshifts. G1 is the brightest galaxy of
the two and is also the closest to the quasar line of sight. In
Figure~\ref{fig:toym}a, we plot the disk halo models for G1 and G2.
The G1 model has velocities that are consistent with up to 150~{\kms}
of the {\MgII} blueward of its systemic velocity. There is a
100~{\kms} gap between the models of G1 and G2. G2 is also
counter-rotating with respect to G1 as viewed from the quasar
sight-line. The G2 model velocities are consistent with the remaining
absorption redward of its systemic velocity.  Although the disk-like
halo model is mostly successful at accounting for some of the
absorption velocity, it does not reproduce all of the {\MgII}
absorption velocities.

\subsubsection{J081420G1}

Galaxy G1 is a low-inclination galaxy with the absorption lining up
exactly with one side of the galaxy rotation curve. In
Figure~\ref{fig:toym}b, we show the halo model velocities and the
{\MgII} absorption profile. Note that there is only a single dashed
line (red) indicating the corrected velocity width since the profile
velocity width is less than the $\pm85$~{\kms} velocity correction. As
we previously mentioned, there is a velocity correction scatter of
$\pm 25$~{\kms} and by taking this into account, the {\MgII}
absorption line is likely a very narrow component centered on the
dashed line.  Regardless, we see that although the absorbing gas is
aligned with the rotation curve, it is in the opposite direction
expected for disk rotation, i.e., the galaxy is ``counter-rotating''
with respect to the {\MgII} absorption.

\subsubsection{J091119G1}

Galaxy G1 is an edge-on spiral with the quasar line-of-sight probing
roughly perpendicular to the galaxy major axis. The {\MgII} $\lambda
2796$ spans both sides of the galaxy systemic velocity. In
Figure~\ref{fig:toym}c the halo model can account for the {\MgII}
absorption blueward of the galaxy systemic velocity. However, the bulk
of the {\MgII} (specifically the $\lambda 2803$ transition) is redward
of the galaxy systemic velocity and is ``counter-rotating'' with
respect to the galaxy's direction of rotation.

\subsubsection{J092300G1}

Galaxy G1 is an S0-like galaxy that does not have emission lines but
exhibits rotation as measured from the absorption lines. Since the
quasar line-of-sight probes along the minor axis of the galaxy, if
halo rotation was responsible for the absorption kinematics, then one
would expect the absorbing gas to have velocities more consistent with
the galaxy systemic velocity. In Figure~\ref{fig:toym}d, we find that
the halo model can adequately account for the total velocity spread
blueward of the galaxy systemic velocity. However, the model can not
explain the $\sim 300$~{\kms} of {\MgII} absorption detected redward
of the galaxy systemic velocity. These high velocities detected along
the quasar line-sight only 12~kpc away only the major axis may be
signatures of outflow or infall. The galaxy does appear to have a
separate optical clump/component or satellite seen below the galaxy
(along the slit) which may be interacting and causing an infall of
metal-enriched gas. We do not have a spectrum of this second object.

\subsubsection{J102847G1}

The strongest component of the {\MgII} absorption associated with G1
aligns exactly with its redward maximum rotation. In
Figure~\ref{fig:toym}e, we find that the galaxy is
``counter-rotating'' with respect with the strongest {\MgII}
component.  In fact, the halo model can not explain the large velocity
width of the absorption. The halo model does not well represent the
kinematics observed along this quasar line-of-sight.

\subsubsection{J111850G1}

The quasar line of sight probes the minor axis of G1 and the
absorption spans the entire rotation velocity. In
Figure~\ref{fig:toym}f, we see that disk rotation can account for the
bulk of the {\MgII} that is redward of the galaxy systemic
velocity. However, it can not account for the absorption blueward of
the galaxy systemic velocity. Thus, this gas is also
``counter-rotating'' with respect to the galaxy and is inconsistent
with disk rotation.

\subsubsection{J114518G1}

The quasar line-of-sight near G1 probes the galaxy major axis with the
bulk of the {\MgII} residing to one side of the galaxy systemic
velocity. In Figure~\ref{fig:toym}g we find that the disk model can
adequately account for the {\MgII} absorption that has velocities
redward of the galaxy systemic velocity along the
line of sight. However, the model does not account for a small
fraction of the {\MgII} absorption blueward of the galaxy systemic
velocity. Thus, the model can account for the majority of the
absorption but can not explain some of the weaker absorption
100~{\kms} blueward of the galaxy systemic velocity.

\subsubsection{J114803G1}

The galaxy G1 exhibits a low level velocity shear. Given the velocity
spread of the {\MgII} absorption, it is impossible for the bulk of the
absorbing gas to be consistent with the observed velocities of G1.
In Figure~\ref{fig:toym}h we see that the galaxy disk halo model is
counter-rotating with respect to the bulk of the absorbing gas. There
is little overlap between the predicted halo model velocities with
those of the {\MgII} absorption. Even if the galaxy had a highly
significant velocity shear, the bulk of the {\MgII} would not be
consistent in velocity space.

\subsubsection{J144033G1 \& J144033G2}

Galaxies G1 and G2 are potentially interacting galaxies, given their
close angular proximity and redshifts. G1 is the brightest galaxy of the two,
however G2 is much closer to the quasar sight-line and is clearly
forming stars. In Figure~\ref{fig:toym}j, we plot the disk halo models
for G1 and G2.  The halo models for G1 and G2 cover roughly the same
velocity range of 100~{\kms} and are rotating in the same direction.
Although the model is mostly successful, it fails to predict the
absorption at higher velocities between $100-200$~{\kms}.

\subsubsection{J161940G1}

The galaxy G1 exhibits a low level velocity shear.  In
Figure~\ref{fig:toym}k, we see that the galaxy disk halo model is
``counter-rotating'' with respect to the dominant saturated {\MgII}
component. There is no overlap between the predicted halo model
velocities with those of the {\MgII} absorption. Even if the galaxy
had a highly significant velocity shear, the bulk of the {\MgII}
clouds would not be consistent in velocity space.

\subsubsection{J225036G1}

The almost edge-on galaxy G1 has the bulk of the {\MgII} residing to
the redward side of the galaxy systemic velocity. In
Figure~\ref{fig:toym}l we see that the disk--halo model can account
for almost all of the {\MgII} absorbing gas velocity spread. It does
not quite account for the gas blueward of the galaxy systemic
velocity, however with the $\pm 25$~{\kms} errors on the corrected
absorption line widths (red lines), we can say that the model is
likely consistent with the absorption velocities. Thus, a disk-like
halo model well represents the absorption kinematics.


\begin{deluxetable*}{lcccrccr}
\tabletypesize{\scriptsize} \tablecaption{Environment Within 0.5~Mpc of
{\MgII} Host Galaxy(ies)\label{tab:env}} \tablecolumns{8}
\tablewidth{0pt}

\tablehead{
\colhead{Galaxy}&
\colhead{Companion} &
\colhead{Companion RA} &
\colhead{Companion DEC} &
\colhead{D} &
\colhead{Mag } &
\colhead{$z_{gal}$} &
\colhead{$\Delta v_{r}$}\\
\colhead{Name}&
\colhead{ID} &
\colhead{(J2000)} &
\colhead{(J2000) } &
\colhead{ (kpc)} &
\colhead{ } &
\colhead{ } &
\colhead{(\kms)}
}
\startdata
J081420G1  &  $\cdots$\phantom{$^a$}  & $\cdots$     &   $\cdots$   &$\cdots$&$\cdots$&  $\cdots$          & $\cdots$  \\   
J102847G1  &  $\cdots$\phantom{$^a$}  & $\cdots$     &   $\cdots$   &$\cdots$&$\cdots$&  $\cdots$          & $\cdots$  \\   
J111850G1  &  $\cdots$\phantom{$^a$}  & $\cdots$     &   $\cdots$   &$\cdots$&$\cdots$&  $\cdots$          & $\cdots$  \\   
J114518G1  &  $\cdots$\phantom{$^a$}  & $\cdots$     &   $\cdots$   &$\cdots$&$\cdots$&  $\cdots$          & $\cdots$  \\   
J114803G1  &  $\cdots$\phantom{$^a$}  & $\cdots$     &   $\cdots$   &$\cdots$&$\cdots$&  $\cdots$          & $\cdots$  \\   
J225036G1  &  $\cdots$\phantom{$^a$}  & $\cdots$     &   $\cdots$   &$\cdots$&$\cdots$&  $\cdots$          & $\cdots$  \\   
J005244G1  &  G2\tablenotemark{a}     & 00:52:44.02  &$-$00:56:46.41& 32    &  19.5   & 0.13465$\pm$0.00002& $+$82.2    \\ 
J091119G1  &  G2                      & 09:11:00.97  & +03:12:33.61 &428    & 16.5    & 0.09705$\pm$0.00005& $-$188.7    \\   
J144033G1  &  G2\tablenotemark{a}     & 14:40:34.56  &$+$04:48:25.10& 25    & 17.2    & 0.11271$\pm$0.00001&  $+$88.8     \\
J161940G1  &  G2                      & 16:19:39.27  & +25:43:33.63 & 383   & 17.2    & 0.12470$\pm$0.00015&  $-$82.7   \\ 
J092300G1  &  G2                      & 09:22:58.65  & +07:52:37.10 & 187   & 17.5    & 0.10368$\pm$0.00015&  $+$149.5   \\ 
           &  G3                      & 09:23:05.55  & +07:47:41.38 & 408   & 17.6    & 0.10315$\pm$0.00016&   $+$293.7   
\enddata 
\tablenotetext{a}{Companion galaxies identified during our
survey and discussed in the text.}
\end{deluxetable*}

\subsection{Summary II: Disk Halo Model}

In an effort to reproduce the observed {\MgII} absorption velocity
spread, we have applied a simple disk halo model to compute the
expected absorption velocities. In only one case, J225036G2, we were
able to reproduce almost the full {\MgII} absorption velocity spread
with the thick disk model.  In four cases, including the two double
galaxy systems, the halo rotation model can account for a large
fraction of the absorption, however, it still can not account for all
of the absorbing gas velocity spread.  In six cases (55\%), the model
is ``counter-rotating'' with respect to the bulk of the {\MgII}
absorption. This indicates that gaseous galaxy halos at $z\sim0.1$ are
likely not co-rotating with their host galaxy.

We emphasize again that our halo model is an extreme case where all of
the gas is assumed to rotate at the maximum observed galaxy rotation
velocity. Under these unrealistic model parameters, the disk halo
model provides insight into the degree at which rotation kinematics
can account for limited regions of the absorption velocity spread.
Relaxing these conditions would significantly diminish the level of
agreement between the model and the observed {\MgII} velocity spread.

The inability for the models to account for all of the halo gas
velocity spread {\it and} direction suggests that additional dynamical
processes are giving rise to some of the {\MgII} absorption (such as
galaxy-galaxy interactions, outflow, infall, or a combination
thereof).

\section{Environment, Outflow, or Infall?}\label{sec:mech}

There are three likely scenarios that could help produce extended
metal-enriched gaseous halos around galaxies: (1) Galaxy group
environments producing tidal streams and stripped gas from galaxies.
(2) Outflowing gas from star-forming regions and/or supernovae, and/or
AGN.  (3) Infalling gas from streams, filaments, high velocity clouds,
satellites and previously ejected gas from outflows.  In this section,
we attempt to determine if the {\MgII} halo gas detected in absorption
is produced via environmental effects, outflow, or infall.

\subsection{Environment}

In Table~\ref{tab:env} we show additional galaxies spectroscopically
identified by SDSS within $\pm 1000$~{\kms} of the {\MgII} absorption
redshift and within 0.5~Mpc (projected) of the quasar sight-line. Six
of the quasar lines-of-sight do not have any near neighbors that can
be further associated with the absorption.  Thus, these six absorption
systems appear to arise within isolated galaxy halos.

The remaining five absorption systems have multiple galaxies at a
range of impact parameters along the quasar sight-lines; two are
galaxy pairs identified during our own spectroscopic surveys
(J005244G1, G2 and J144033G1, G2). Other than these two less massive
satellite galaxies identified here (G2s), there are no other galaxies
associated with these absorbers within the limits defined above. For
both of these absorbers, we have shown that the galaxies are within
the standard {\MgII} halo size, and their galaxy dynamics are
consistent with the absorption velocities. It is possible that, given
their close proximity in both projected distance and line-of-sight
velocity, they may have undergone some interactions in the past.
Although, given the luminosity ratios for these galaxies pairs, both
companion galaxies are more consistent with being satellite galaxies
within a main galaxy halo.

The remain 3/5 absorption systems that have multiple galaxies within
$\pm 300$~{\kms} of the {\MgII} absorption redshift tend to be at
much higher impact parameters. The impact parameters range between
190--430~kpc: much larger than the standard halo size. Thus, for the
most part, these galaxies would appear similar to isolated galaxies.

To further study the environments of these $z=0.1$ {\MgII} absorbing
and non-absorbing galaxies, \citet{barton09} created an artificial
redshift survey through cosmological dark matter simulations.  They
found that {\MgII} host galaxies appear to reside low-density
environments while non-absorbing galaxies seem to reside denser
regions and are likely to have companions. They further suggest that
non-absorbing galaxies may result from stripping of the outer gas halo
as galaxies fall into a denser environment. This truncation of halo
sizes may decrease by as much as a factor of 10 when observed in
clusters \citep{padilla09}.  Given that we detected {\MgII} absorption
over a large range of impact parameters and that the host galaxies
appear to be in isolated, galaxy environment does not appear to be a
strong factor in determining the absorption strength of the halo gas.

In our sample, environment may not play a crucial role in producing
the extended metal enriched galaxy halos. This result is supported by
the data as well as mock surveys through cosmological simulations.
Although, two of our systems hint that minor interactions occur and
likely produce a combination of streams plus an additional source of
inflowing gas towards the host galaxy.

\subsection{Outflows}

Here we explore evidence of outflows using two techniques: we first
compute the host galaxy SFRs and compare them to the halo gas
absorption strength. We then compute {\MgIb} (stellar) and {\NaID}
(stellar+ISM) line ratios in order to identify possible outflows.

\begin{figure}
\begin{center}
\includegraphics[scale=0.52]{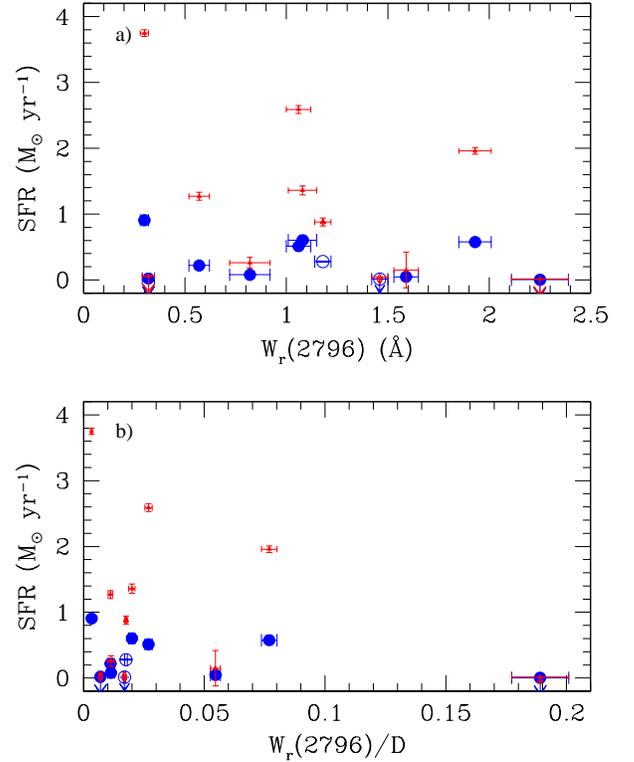}
\caption{--- (a) SDSS fiber SFR computed for the absorbing galaxies as
a function of $W_r(2796)$. The open circles are the two galaxies
J005244G1 and J144033G1 that have minor companions. The small (red)
triangles are the SDSS aperture corrected SFRs. Note there is no tread
with SFR and galaxies with little star formation are associated with
some of the high equivalent width systems. --- (b) The {\MgII}
equivalent width normalized by impact parameter as a function of
SFR. Again, we find no correlation suggest an absence of star
formation driven winds. }
\label{fig:SFR}
\end{center}
\end{figure}

\subsubsection{Star Formation Rates}

We have computed star formation rates (SFRs) for the host galaxies
where possible. We note that the SDSS spectra are obtained using
fibers that have an aperture radius of 1.5$''$, which translates to
2.77~kpc at $z=0.10$.  Our galaxies have an average half-light radius,
as measured from GIM2D (see Table~\ref{tab:params}), of $<r_h>=5.6\pm
1.3$~kpc, so the SDSS fibers cover only the inner regions of the
galaxies. Therefore, SFRs computed here are only within the
fiber. This still provides useful information since all galaxies are
roughly at the same redshift and therefore probing the same physical
scales, and strong winds are expected to originate within the central
regions.

For the fiber SFRs we have applied Galactic extinction correction
obtained from NED \footnote{http://nedwww.ipac.caltech.edu/}. We are
not able to apply Balmer decrement corrections since only two galaxies
have detected {\Hb} emission. The {\Ha} luminosities were measured
from the SDSS spectra and {\Ha}-derived SFRs were computed using the
formalism of \citet{kewley02}.We also performed aperture loss
corrections to the SFRs. The applied scaling factor was determined
from the ratio of the $r-$band galaxy total counts to those within the
SDSS fiber.

In Figure~\ref{fig:SFR}a, we show the inner galaxy (fiber) SFRs, and
corrected SFRs, as a function of the {\MgII} absorption strength. We
find no correlation between the SFRs and the halo gas absorption
strength. A similar distribution arises if one plots {\Ha} against
$W_r(2796)$.  Note that some galaxies have only low SFR limits yet are
still associated with strong absorption.  Given that \citet{barton09}
noted that red galaxies appear to be closer to the quasar
line-of-sight then blue star-forming galaxies, we normalized out the
impact parameter in Figure~\ref{fig:SFR}b: we find no statistically
significant trend here.

Another indicator of galaxy outflows is the star formation per unit
area.  \citet{heckman02,heckman03} demonstrated that outflows are
ubiquitous in galaxies where the global SFR per unit area exceeds
$\Sigma = 0.1$~M$_{\odot}$~yr$^{-1}$~kpc$^2$ (the area is defined by
the half-light radius).  The ISM entrained in these winds have outflow
speeds of $\sim$100 to $\sim$1000~{\kms}. Although the half-light
radii of our galaxies are larger than the SDSS fibers, we can compute
the surface star formation density within the SDSS fiber. For our
sample, we find that the star formation per unit area of $0.03 \leq
\Sigma \leq 0.0002$~M$_{\odot}$~yr$^{-1}$~kpc$^2$, which is well below
what is expected for strong winds.

These results possibly suggest that star-formation driven winds, at
least in the galaxy central regions, are not producing the observed
kinematics and absorption strength of the metal enriched halo gas. It
is possible that the metal enriched gas detected along the quasar
sight-lines are potential reservoirs for future star formation.

\begin{figure}
\begin{center}
\includegraphics[scale=0.43]{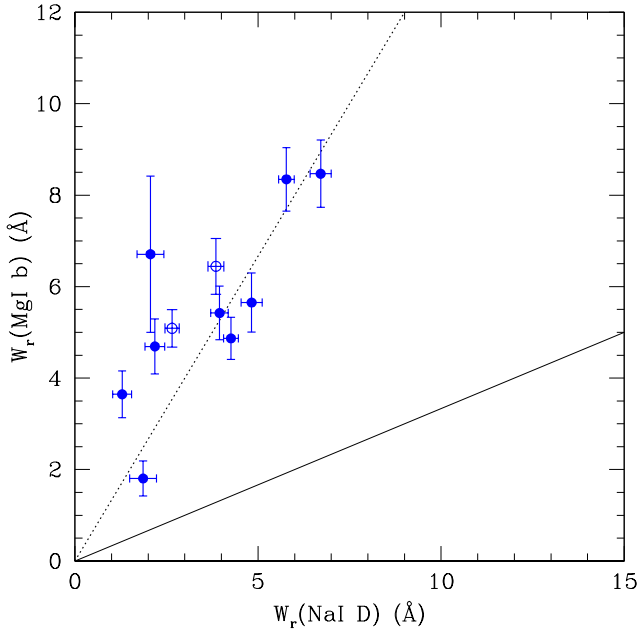}
\caption{ The equivalent widths of the {\NaID} (stellar+ISM) and
  {\MgIb} (stellar) absorption lines. The open circles are the two
  galaxies J005244G1 and J144033G1 that have minor companions.  The
  lower solid line is the relation of \citet{rupke05a,rupke05b} who
  found that 80\% of galaxies below this line have winds, while only a
  small fraction above the solid line have winds (25\%). The dotted
  line represents the expected stellar contribution to the {\NaID} by
  scaling the equivalent width of {\MgIb}. Note that our galaxies
  reside near this relation and far from where winds are expected to
  dominate.}
\label{fig:winds}
\end{center}
\end{figure}

\subsubsection{{\NaID} and {\MgIb} Line Ratios}

It has been demonstrated that {\NaID} and {\MgIb} absorption line
ratios are good tracers of outflows. Both {\NaID} and {\MgIb} have
similar ionization potentials (5.14 eV for {\NaID}, 7.65 eV for
{\MgIb}). Although both {\NaID} and {\MgIb} appear in stellar spectra,
where the absorption strength peaks in spectra of cool K--M stars
\citep[see][]{jacoby84}, the {\MgIb} band is a highly excited
transition making it purely stellar in origin.  On the other hand, the
{\NaID} resonance line can also be absorbed by the ISM and has been
detected as entrained gas within galactic scale winds
\citep[e.g.,][]{martin05,rupke05a,rupke05b,heckman00}. Thus, this line
ratio can be used to successfully separate starburst outflowing galaxies
from quiescent galaxies with little-to-no winds
\citep[e.g.,][]{heckman00}.

We have used the {\NaID} and {\MgIb} equivalent widths computed by
SDSS, as observed in galaxy spectra to study the wind properties of
our galaxy sample.  Again, we note that the SDSS spectral fibers have
an aperture radius of 1.5$''$, which translates to 2.77~kpc at
$z=0.10$, thus the fibers only cover the central regions of the
galaxies. The line ratio is only computed within the fiber, however,
galactic winds are expected to originate from the centers of galaxies,
which is where the wind-signature in the line ratio is expected.  As
one would include more of the galaxy light, from regions where no
winds are found, we would expect the wind-line ratios to be more
consistent with stellar origins.

In Figure~\ref{fig:winds}, we show the {\NaID} and {\MgIb} equivalent
width distribution.  \citet{heckman00} estimated the expected stellar
contribution to the {\NaID}, by scaling the equivalent width of
{\MgIb}, and found the relation $W_r($\NaID$) = 0.75 W_r($\MgIb$)$
represented by the dotted line in Figure~\ref{fig:winds}.  Galaxies
that reside to the right of the solid line are most likely to have
contributions from the ISM to the {\NaID} absorption.  The solid line,
which is expressed as $W_r($\NaID$) = 3 W_r($\MgIb$)$, is the
approximate location of starbursting galaxies with strong {\NaID}
winds observed by \citet{rupke05a,rupke05b}. They found that 80\% of
galaxies below this line have winds, while only a small fraction above
the solid line have winds (25\%).

All of our galaxies reside far from the relation where winds are
expected and reside tightly near the expected stellar contribution
line, suggesting that our galaxy sample have little to no strong
winds. Scatter about that line is most likely due to interstellar
{\NaID} absorption.

Another way to detect outflows is to observe velocity offsets between
galaxy nebular emission lines and absorption lines. This technique has
been applied in previous studies on a range of absorption lines and
has been demonstrated to detect strong winds
\citep[see][etc.]{heckman00,rupke05b,weiner09,steidel10,rubin10}. Eight
of our absorbing galaxies have measurable {\Ha} and {\NaID} lines.
Although we have discussed above that {\NaID} absorption is
contaminated by stellar absorption, it still can be used to trace
winds from velocity offsets from the emission lines
\citep[e.g.,][]{heckman00,rupke05b}.  Furthermore, \citet{heckman00}
found that galaxies with {\NaID} absorption residing within
$\pm$70~{\kms} of the systematic galaxy redshift were consistent with
a predominantly stellar origin. {\NaID} absorption blue-shifted with
velocities greater than 100~{\kms} were associated with outflows
$\sim70\%$ of the time.  They also found that galaxies with strong
outflows were viewed at low inclination angles.

For the eight galaxies in our sample, we find a mean velocity
intrinsic {\NaID} absorption offset of $\Delta v=-71\pm 26$~{\kms},
consistent with a predominantly stellar origin. However, all galaxies
have a negative velocity offset from the {\Ha} emission line. This
offset may suggest low level winds, although, we find no correlation
between the velocity offset and the galaxy SFR, or $\Sigma$, as would
be expected \citep[][]{weiner09}.

The average galaxy inclination angle is 46 degrees which is less than
the average expected for a random distribution of galaxies. However,
we do not find a significant correlation between the velocity offset
and the galaxy inclination or absorption strength.

These results suggest that active outflows are not responsible for the
dominant component of the {\NaID} absorption and are consistent with a
stellar component plus some contribution from dynamically cold
interstellar gas. The lack of strong outflows close to the galaxy
suggests alternate origins of the {\MgII} halo gas ($0.3 \leq
W_r(2796) \leq 2.3$~\AA).

\section{Discussion}\label{sec:discussionkine}

\citet{kacprzak10a} compared {\MgII} absorption and galaxy rotation
kinematics of 10 $W_r(2796) < 1.4$~{\AA} systems and found that, in
most cases, the absorption was fully to one side of the galaxy
systemic velocity and usually aligned with one arm of the rotation
curve.  These results are consistent with earlier studies of five
galaxies by \citet{steidel02}, one galaxy by \citet{ellison03}, and
three galaxies of \citet{chen05}.  In only 5/19 cases, the absorption
velocities spans both sides of the galaxy systemic velocity. Three of
those have $W_r(2796)>1$~{\AA} and their absorption kinematics
displayed possible signatures of winds or superbubbles
\citep{bond01a,ellison03}; two of these galaxies have SFRs and SFRs
per unit area consistent with wind-dominated galaxies
\citep{kacprzak10a}.

For our $z\sim 0.1$ sample, we find that for only 5/13 galaxies the
{\MgII} absorption resides to one side of the galaxy systemic velocity
and aligns with one side of the rotation curve. For the remaining 8/13
galaxies, the absorption spans both sides, although the bulk of the
{\MgII} resides mostly to one side of the galaxy systemic velocity.

In comparing the results from $z\sim 0.5$ to our $z\sim 0.1$ sample,
we find a factor of three increase in the fraction of systems where
the {\MgII} absorption resides on both sides of the galaxy systemic
velocity at lower redshift. These results may suggest an evolution in
the halo gas kinematics as a function of redshift. Both sample span a
similar range of impact parameters and {\MgII} equivalent width.
Hints that halo gas properties may evolve with redshift have already
been observed.  \citet{barton09} found that gas covering fractions may
decrease by a factor of $2-3$ by $z=0.1$.  It is also important to
note that the sample of \citet{kacprzak10a} and \citet{steidel02} have
an average $<L_B> = 0.6L^*_B$ whereas our sample has a $<L_B> =
1.9L^*_B$: in this study we have probed more massive galaxies at lower
redshift.  This leads to the possibility that there could be an
evolution as a function of host galaxy mass.  This is consistent with
the cosmological SPH simulations of \citet{stewart10} who predicts
that the halo gas covering fraction exhibits a sharp decrease when the
galaxy mass exceeds a critical minimum mass to form stable shocks
which results in a transition from cold mode to hot mode gas
accretion. This can reduce the covering fraction, over the same
redshifts observed here, by a factors of $6-10$.  It has also been
demonstrated that for massive galaxy halos of $\gtrsim
10^{13}$h$^{-1}$~M$_{\odot}$ at $z\sim0.5$, the covering fractions
decrease by a factor of $7-15$ \citep{gauthier10,bowen10}. A more
uniform sample is required to explore the possibility of an evolution
as a function of mass and/or redshift.

Since the halo gas velocities at intermediate redshift were found to
align in the same sense and as velocities expected for co-rotation, it
strongly suggest ``disk-like'' rotation of the halo gas. Both
\citet{kacprzak10a} and \citet{steidel02} applied simple co-rotating
disk halo models and concluded that an extension of the disk rotation
was able to explain some of the gas kinematics. However, the models
were not able to account for the full absorption velocity spreads.

Here we obtain a similar conclusion, except for one case, all of the
observed kinematics velocity spread of the halo gas can not be
explained with a simple rotating disk model.  However, contrary to
previous studies, for 55\% of our sample, the halo model is
``counter-rotating'' with respect to the bulk of the {\MgII}
absorption, and in two cases, there is zero overlap between the model
and the absorption velocities.  This suggests that at least some
gaseous halos at $z\sim0.1$ are not co-rotating with their host
galaxies {\it and} to a lesser extent than what was found at $z\sim
0.5$. Again this implies that other mechanisms must be invoked to account
for the full velocity spreads.

In an effort to identify the origins of the {\MgII} absorption,
\citet{kacprzak10a} used hydrodynamical cosmological simulations,
combined with the quasar absorption line methods, to demonstrate that
the majority of the {\MgII} absorption arises in an array of
cosmological structures, such as filaments and tidal streams. They
showed that metal-enriched gas was infalling towards the galaxy along
these structures with velocities in the range of the rotation velocity
of the simulated galaxy and consistent with the observed galaxy halo
gas kinematics. In this paper, we have not gone to these efforts to
explore the origins of the halo gas, however this will be part of our
future work. We have instead chosen to explore the galaxy environment
and also physical properties of the host galaxies that are indicative
of strong outflows.

The $z\sim 0.1$ {\MgII} galaxies appear to be isolated, aside from the
two double galaxy systems identified in our own survey. Again, these
two pairs are consistent with one dominate host galaxy and a smaller
satellite galaxy. Only three host {\MgII} absorbing galaxies have
other nearby galaxies, however they reside far from the quasar
line-of-sight making it unlikely that they contribute to the absorbing
gas. Thus, for our sample, interactions may not play a crucial role in
producing the halo gas absorption.

It is well known that highly star-forming galaxies tend to have strong
outflows that are also detected in absorption
\citep[e.g.,][]{weiner09,martin09,nestor10}. However, for our sample
the host galaxy SFRs computed within the SDSS fiber, representing the
galaxy central regions, are all less than 1~M$_{\odot}$~yr$^{-1}$. In
addition, we find no correlation between the SFRs and the $W_r(2796)$
even when normalized by the impact parameter.
\citet{heckman02,heckman03} demonstrated that outflows with speeds of
$\sim$100 to $\sim$1000~{\kms} are ubiquitous in galaxies where the
global SFR per unit area exceeds $\Sigma =
0.1$~M$_{\odot}$~yr$^{-1}$~kpc$^2$.  For our sample we find $\Sigma
\leq 0.03$~M$_{\odot}$~yr$^{-1}$~kpc$^2$, which is well below what is
expected for strong winds.

We also find that the {\NaID} (stellar$+$ISM) and {\MgIb} (stellar)
absorption line ratios are consistent with being predominately stellar
in origin and having kinematically cool ISM.  The velocity offsets
between the {\NaID} line and the nebular {\Ha} emission line are on
average $-71\pm 26$~{\kms}. Although the shift is in the negative
direction, which is associated with outflows, the velocity shifts are
small and consistent with little-to-no outflows \citep{heckman00}. In
addition, the velocity offsets between {\NaID} and {\Ha} do not
correlate with SFR or $\Sigma$.

Our sample of galaxies appear to be isolated, and undergoing some star
formation, but too little to be producing strong outflows. We find it
is unlikely that the {\MgII} gas originates from either environmental
effects, such as galaxy-galaxy interactions or mergers, and/or
outflowing gas; although accretion of cold gas ejected from previous
star formation events is possible. We favor a scenario where the metal
enriched halo gas is infalling onto the host galaxy with velocities
comparable to the dynamics of their host.

\section{Conclusions}
\label{sec:conclusionkine}

We have examined and compared the detailed galaxy and {\MgII}
absorption kinematics for a sample of 13 intermediate redshift, $\sim
L_{\star}$ galaxies along 11 quasar sight-lines. The galaxy--quasar
impact parameters range from $12 \leq D \leq 90$~kpc.  The galaxy
rotation curves were obtained from DIS/APO and ESI/Keck spectra and
the {\MgII} absorption profiles were obtained from LRIS/Keck quasar
spectra. In an effort to compare the relative kinematics, we used a
disk halo model to compute the expected absorption velocities through
a monolithic gaseous halo model.  We further examined the host galaxy
environments and also studied the intrinsic host galaxy properties,
using them to quantify and identify strong outflows.

Our mains results can be summarized as follows:

\begin{enumerate}

\item For all 13 galaxies, the velocity of the strongest {\MgII}
  absorption component lies in the range of the observed galaxy
  rotation curve.  We find that for the nine isolated galaxy/absorber
  pairs, seven galaxies have well defined rotation curves while two
  galaxies display only shear. In 3/9 cases the absorption resides to
  one side of the galaxy systemic velocity and the absorption redshift
  tends to align with one side of the rotation curve. In the remaining
  6/9 cases, the absorption spans both sides of the galaxy systemic
  velocity, although the bulk of the {\MgII} resides mostly to one
  side of the galaxy systemic velocity.

  For our double galaxy/absorber pairs, we find that all four galaxies
  exhibit well-defined rotation curves. In one case, the absorbing gas
  spans both sides of both host galaxy systemic velocities. In the
  other case, the absorbing gas resides to one side of the systemic
  velocity of both absorbing galaxies.

\item In all cases, the thick disk rotating halo models are unable
  reproduce the full spread of observed {\MgII} absorption velocities.
  Contrary to previous studies at higher redshifts, for 55\% of our
  sample the halo model is ``counter-rotating'' with respect to the
  bulk of the {\MgII} absorption and in two cases there is zero
  overlap between the model and the absorption velocities. This
  potentially suggests that $z\sim0.1$ gaseous halos are not
  co-rotating with their host galaxies {\it and} to a lesser extent
  than what was found at $z\sim 0.5$.  In this simple scenario, even
  if some of the absorbing gas arises in a thick disk, there must be
  dynamical processes (such as infall, outflow, supernovae winds,
  etc.)  that give rise to the remaining {\MgII} absorption.

\item Host galaxy SFRs are all $\lesssim 1$~M$_{\odot}$~yr$^{-1}$ and
  we find no correlation between the SFRs and the $W_r(2796)$ even
  when normalized by the impact parameter. Their SFRs per unit area
  are $\leq0.03$~M$_{\odot}$~yr$^{-1}$~kpc$^2$, which is well below
  the lower limit of 0.1~M$_{\odot}$~yr$^{-1}$~kpc$^2$ expected for
  strong winds.

\item We find our absorbing galaxies tend to be isolated, or at least
  in low density environments. This is further supported by an
  analysis of cosmological simulations at $z\sim 0.1$ performed by
  \citet{barton09}.

\item The {\NaID} (stellar$+$ISM) and {\MgIb} (stellar) absorption
  line ratios are consistent with being predominately stellar in
  origin and having kinematically cool ISM.  The velocity offsets
  between the {\NaID} line and the nebular {\Ha} emission line are on
  average $-71$~{\kms}. Although the shift is in the negative
  direction the velocity shifts are small and are not correlated with
  SFR or $\Sigma$. These results are consistent with our {\MgII}
  absorption-selected galaxies having little-to-no outflows.

\end{enumerate}

In our detailed study of 13 $z=0.1$ absorbers, we find it unlikely
that the {\MgII} gas originates from either outflowing gas and/or
environmental effects. These results are consistent with the
hydrodynamical simulations of \citet{kacprzak10a} where the {\MgII}
gas is inflowing along the streams and filaments with kinematics
comparable to the host galaxy.  Thus, we favor a scenario of infalling
gas that provides a gas reservoir for star formation at these low
redshifts.

\acknowledgments 

We thank Michael Murphy for advice and for carefully reading
manuscript and providing useful comments.  CWC was supported from NSF
grant AST-0708210. EJB gratefully acknowledges support from NSF grant
AST-1009999 and the UC Irvine Center for Cosmology. JC gratefully
acknowledges generous support by the Gary McCue Postdoctoral
Fellowship.  This work is based on observations obtained with the
Apache Point Observatory 3.5-meter telescope, which is owned and
operated by the Astrophysical Research Consortium.  Observations were
also obtained at the W.M. Keck Observatory, which is operated as a
scientific partnership among the California Institute of Technology,
the University of California and the National Aeronautics and Space
Administration. The Observatory was made possible by the generous
financial support of the W.M. Keck Foundation. This paper would not
have been possible without data from The Sloan Digital Sky Survey
(SDSS).  Funding for the SDSS and SDSS-II has been provided by the
Alfred P. Sloan Foundation, the Participating Institutions, the
National Science Foundation, the U.S. Department of Energy, the
National Aeronautics and Space Administration, the Japanese
Monbukagakusho, the Max Planck Society, and the Higher Education
Funding Council for England. The SDSS Web Site is
http://www.sdss.org/. The SDSS is managed by the Astrophysical
Research Consortium for the Participating Institutions. The
Participating Institutions are the American Museum of Natural History,
Astrophysical Institute Potsdam, University of Basel, University of
Cambridge, Case Western Reserve University, University of Chicago,
Drexel University, Fermilab, the Institute for Advanced Study, the
Japan Participation Group, Johns Hopkins University, the Joint
Institute for Nuclear Astrophysics, the Kavli Institute for Particle
Astrophysics and Cosmology, the Korean Scientist Group, the Chinese
Academy of Sciences (LAMOST), Los Alamos National Laboratory, the
Max-Planck-Institute for Astronomy (MPIA), the Max-Planck-Institute
for Astrophysics (MPA), New Mexico State University, Ohio State
University, University of Pittsburgh, University of Portsmouth,
Princeton University, the United States Naval Observatory, and the
University of Washington.





{\it Facilities:} \facility{ARC (DIS)}, \facility{Keck II (ESI)},
\facility{Keck I (LRIS)}, \facility{Sloan (SDSS)}.

\end{document}